\documentclass[aps,floatfix,twocolumn]{revtex4-1}
\usepackage{amsmath}
\usepackage{amsfonts}
\usepackage{amssymb}
\usepackage{graphics,epsfig}
\usepackage{graphicx}
\usepackage[usenames]{color}
\begin{document}

\title{The Vector Volume and Black Holes}
\author{William~Ballik}
\email[]{bballik@astro.queensu.ca}
\author{Kayll Lake}
\email[]{lake@astro.queensu.ca}
\affiliation{Department of Physics, Queen's University, Kingston,
Ontario, Canada, K7L 3N6 }
\date{\today}

\begin{abstract}
By examining the rate of growth of an invariant volume $\mathcal V$ of some spacetime region along a divergence-free vector field $v^\alpha$, we introduce the concept of a ``vector volume" $\mathcal{V}_v$. This volume can be defined in various equivalent ways.  For example, it can be given as $\mathrm d \mathcal V(\mu) / \mathrm d \mu$, where  $v^\alpha \partial_\alpha = \mathrm d / \mathrm d \mu$, and $\mu$ is a parameter distance  along the integral curve of $v$. Equivalently, it can be defined as $\int v^\alpha \mathrm d \Sigma_\alpha$, where $\mathrm d \Sigma_\alpha$ is the directed surface element.  We find that this volume is especially useful for the description of black holes, but it can be used in other contexts as well. Moreover, this volume has several properties of interest.  Among these is the fact that the vector volume is linear with respect to the the choice of vector $v^\alpha$.   As a result, for example, in stationary axially symmetric spacetimes with timelike Killing vectors $t^\alpha$ and axial symmetric Killing vectors $\phi^\alpha$, the vector volume of an axially symmetric region with respect to the vector $t^\alpha + \Omega \phi^\alpha$ is equal for any value of $\Omega$, a consequence of the additional result that $\phi^\alpha$ does not contribute to $\mathcal{V}_v$. Perhaps of most interest is the fact that in Kerr-Schild spacetimes the volume element for the full spacetime is equal to that of the background spacetime. We discuss different ways of using the vector volume to define volumes for black holes.  Finally, we relate our work to the recent wide-spread thermodynamically motivated study of the ``volumes" of black holes associated with non-zero values of the cosmological constant $\Lambda$.
\end{abstract}
\maketitle
\section{Introduction}

It is well known (e.g.~\cite{Poisson}) that, following the pioneering work of Bardeen, Carter and Hawking \cite{Hawking}, the ``surface area'' of black holes ($\mathcal{A}$) is of fundamental interest.  It is non-decreasing through classical (non-quantum) processes and is thus generally associated with the black hole entropy.  This raises the naive question as to whether or not the ``volume" of black holes is important. This question has seen little interest until recently. With the surge in interest in the cosmological constant $\Lambda$, since $\Lambda$ is, naively, a ``pressure" term, one can certainly ask where the ``$PdV$" term went in the first law of black hole thermodynamics. There is now wide-spread interest in this ``$V$" term, e.g.~\cite{CaldarelliEtal:2000, KastorEtal:2009, Dolan:2010, Dolan:2011a, Dolan:2011b, Dolan:2012, Cvetic, LarranagaCardenas:2012, LarranagaMojica:2012,Gibbons:2012, KubiznakMann:2012, GunasekaranEtal:2012, BelhajEtal:2012,  LuEtal:2012, SmailagicSpallucci:2012, HendiVahinidia:2012, dolan}.  However, while the surface area is well-defined and invariant for a $D$-dimensional black hole, the volume intrinsic to a $(D-1)$-dimensional hypersurface depends on the choice of slicing, and the full $D$-dimensional volume of the black hole region is formally infinite.

The attempt to find a definition of volume which is both invariant and finite is the inspiration for this work in which we define a ``vector volume''. This can be thought of as the rate of change of the invariant four-volume of some region along a vector field.  The vector volume turns out to be a generalization of several other definitions of volume in use by other authors.  Moreover, the vector volume has many interesting properties which we examine here.  We show, for example, that in certain ``Kerr-Schild" cases, the vector volume reduces to the Euclidean volume of a subspace region. We also offer a new definition of the surface gravity as the ratio of two vector volumes.

The paper is structured as follows: In Section \ref{review}, we review recent developments in which various authors working independently have used similar expressions for black hole volume.  In Section \ref{vecvolintrodef} we define the vector volume in several different ways which we show to be equivalent.  In Section \ref{generalproperties} we show some of the vector volume's interesting properties.  Section \ref{kerrschildtype} discusses the vector volume in Kerr-Schild-type metrics.  Section \ref{canon} discusses a ``canonical black hole volume'' based on the vector volume using the time-translation Killing vector in stationary black holes which is closely related to work from Parikh \cite{Parikh} and Cveti\v{c} et al.~\cite{Cvetic}  Section \ref{ourvolume} discusses a ``null generator volume'' for stationary black holes, first defined in \cite{arXivPaper}, and how this can be used alongside the canonical black hole volume to define the surface gravity of black holes in a novel way.  Section \ref{otherauthors} further explores the connection between the vector volume and other works, concentrating on those of Cveti\v{c} et al.~\cite{Cvetic} and Hayward \cite{Hayward}.  Section \ref{conclusion} summarizes our analysis.

We work largely in dimension $D = 4$ for clarity, though in some cases we keep the dimension $D$ of the spacetime general.~Most of what follows is generalizable to higher dimensions though this generalization is not our principal concern here.

\section{Review of Some Recent Developments} \label{review}

Here we review several independent recent volume definitions for black holes.  It turns out that these are all special cases of a more general vector volume which we present in this paper.  The work is from Parikh \cite{Parikh}, Cveti\v{c} et al.~\cite {Cvetic}, and Hayward (e.g.~\cite{Hayward}).  For the most part we use the notation used by the authors in their papers.  We note that the authors of this paper have independently developed a definition of volume which is in a similar vein to these \cite{arXivPaper}.

\subsection{Parikh Volume} \label{Parikhintro}

In 2006, Parikh \cite{Parikh} defined a volume for stationary black holes.  In his paper, Parikh begins by considering a $D$-dimensional spherically symmetric spacetime with a timelike Killing vector and a horizon, with line element \cite{notation}
\begin{equation}\label{metricone}
\mathrm d s^2 = -\alpha(r) \mathrm d t_s^2 + \frac{\mathrm d r^2}{\alpha(r)} + r^2 \mathrm d \Omega_{D-2}^2(\vec x)
\end{equation}
where $\mathrm d \Omega_{D-2}$ is the line element for the $(D-2)$-sphere, $r$ is the aerial radius, $\vec x$ is a vector representation of the coordinates on the $(D-2)$-sphere, $t_s$ is the static time coordinate and $\alpha(r)$ is some function of $r$.  The Killing vector $\partial_{t_s}$ is the Killing vector corresponding to staticity, which is timelike outside the horizon.  The horizon is at a radius $r = r_+$ for which $\alpha(r) = 0$.  At this value, the metric is non-regular, so Parikh introduces a new time coordinate $t$ defined by
\begin{equation}
t_s = \lambda t + f(r,\vec x).
\end{equation}
$\partial_t$ will be a Killing vector for any constant $\lambda$, but in order to preserve the time orientation and asymptotic normalization of the Killing vector, $\lambda$ is set to +1 in Parikh's definition, so that $\partial_t = \partial_{t_s}$.  The advantage to this new coordinate $t$ over $t_s$ is that for certain functions $f$ it is possible to obtain a slicing that extends through the horizon.

Parikh then notes that while the $(D-1)$-dimensional volume of the region $0 \leq r \leq r_+$ on the hypersurfaces of constant $t$ depends on the choice of function $f$, one can instead define a ``differential spacetime volume'' $\mathrm d \mathcal V_D$ which is invariant.  This differential spacetime volume is the $D$-volume of the region where $t'$ varies between $t$ and $t + \mathrm d t$:
\begin{equation}
\mathrm d \mathcal V_D = \int_t^{t + \mathrm d t} \mathrm d t' \int_0^{r_+} \mathrm d r \int d^{D-2}x \sqrt{- g_D}.
\end{equation}
Here, $g_D$ is the determinant of the full $D$-dimensional metric.  Since $\partial_t$ is a Killing vector, the metric $g_{\alpha \beta}$ is independent of $t$, and thus $t$ enters into $\mathrm d \mathcal V_D$ only through the multiplicative term $\mathrm d t$. The Parikh volume is defined as the ratio of this differential spacetime volume to $\mathrm d t$:
\begin{equation}
\mathcal V_{P} \equiv \frac{\mathrm d \mathcal V_D}{\mathrm d t} = \int \mathrm d^{D-1}x \sqrt{-g_D} \label{Parikhdef}
\end{equation}
where $\mathrm d^{D-1}x$ is the product of the differentials except for $\mathrm d t$.  Essentially, one uses $g_D$ instead of $g_{D-1}$ (the determinant of the metric of the $t = const.$ hypersurfaces), which makes the volume constant in time for all choices of Killing time, and invariant under the choice of stationary time slices.

Though he used static spherical symmetry as an example, Parikh notes that his volume definition can similarly be applied to any stationary black hole. In particular, Parikh notes that the volume for static, spherically symmetric black holes in 4 dimensions, as before with horizon at $r = r_+$, is
\begin{equation}
\mathcal V_{P} = \frac{4\pi}{3} r_+^3,
\end{equation}
which is of course simply the Euclidean volume for a sphere of radius $r_+$. The volume for the (four-dimensional) Kerr black hole is given by Parikh as
\begin{equation} \label{vpk}
\mathcal V_{P} = \frac{4\pi}{3} r_+(r_+^2 + a^2)
\end{equation}
where $r_+$ and $a$ have their usual meanings as the value of radius $r$ at the outer horizon and specific angular momentum respectively.

Before continuing, we make a further note about the Parikh volume.  Using the well-known horizon area $\mathcal{A}=4 \pi (r_{+}^2+a^2)$ we can write (\ref{vpk}) in the form
\begin{equation} \label{vpa}
\mathcal V_{P} = \frac{r_{+} \mathcal{A}}{3},
\end{equation}
a result we revisit in future sections.

\subsection{Geometric Volume} \label{cveticgeometricvolume}

The laws of black hole thermodynamics with a non-zero cosmological constant term $\Lambda$, as well as the generalized Smarr formula, where put into a general geometrical approach in \cite{KastorEtal:2009}. Here, to be specific, we follow \cite{Cvetic}. (See as well the references therein and subsequent work \cite{LarranagaCardenas:2012, LarranagaMojica:2012,Gibbons:2012, KubiznakMann:2012, GunasekaranEtal:2012, BelhajEtal:2012,  LuEtal:2012, SmailagicSpallucci:2012, HendiVahinidia:2012, dolan}.)  The argument goes in essence as follows.  In black hole spacetimes with $\Lambda$, the black hole thermodynamic variation laws can be written in terms of a black hole enthalpy $E$, giving rise to a modified first law of thermodynamics,
\begin{equation}
\mathrm d E = T \mathrm d S + \sum_i \Omega_i \mathrm d J_i + \sum_\alpha \Phi_\alpha \mathrm d Q_\alpha + \Theta \mathrm d \Lambda,
\end{equation}
or, in non-differential Smarr-Gibbs-Duhem form,
\begin{equation}\label{sgd}
E = \frac{D-2}{D-3}\left(T S + \sum_i \Omega_i J_i\right) + \sum_\alpha \Phi_\alpha Q_\alpha - \frac{2}{D-3} \Theta \Lambda,
\end{equation}
where $T$ is the effective temperature of the black hole, $S$ is the entropy, $J_i$ are the components of the angular momenta, $\Omega_i$ are the corresponding angular velocities, $Q_\alpha$ are the conserved charges, $\Phi_\alpha$ are the potentials corresponding to those charges, and $\Theta$ is the conjugate to $\Lambda$.  Since $\Lambda$ can be interpreted as a pressure (up to a multiplicative constant), $\Theta$ is interpreted as being proportional to a volume for the black hole, by analogy with the classical thermodynamical first law for enthalpy $H$ in terms of temperature $T$, entropy $S$, pressure $P$, volume $V$ and work $W$,
\begin{equation}
\mathrm d H = T \mathrm d S  - \delta W + V \mathrm d P.
\end{equation}
This yields a relationship between a ``thermodynamic" volume $\mathcal V_{th}$ and $\Theta$:
\begin{equation}
\mathcal V_{th} = - \frac{16 \pi \Theta}{D-2},
\end{equation}
where $D$ is the dimension of the spacetime. With spherical symmetry, $\mathcal V_{th}$ corresponds to the ``naive" geometrical volume
\begin{equation}
\mathcal V_{geo} = \int \mathrm d r \int \mathrm d \Omega \sqrt{- g_D} \label{Vgeo}
\end{equation}
where $r$ ranges over the black hole and $\mathrm d \Omega$ is the surface element on the unit $(D-2)$ sphere. For black holes with non-zero angular momentum, the thermodynamic and gemoetric volumes differ by \cite{Cvetic}
\begin{equation}\label{vthermo}
\mathcal V_{th} - \mathcal V_{geo} =  \frac{8 \pi}{(D-1)(D-2)} \sum_i a_i J_i,
\end{equation}
where $a_i$ are the rotational parameters for the black hole corresponding to the $J_i$.  The geometric volume satisfies the relation
\begin{equation}
\mathcal V_{geo} = \frac{r_+ \mathcal{A}}{D-1} \label{VrA1}
\end{equation}
for all black holes of the Kerr-Newman-de Sitter family, generalizing (\ref{vpa}). Again, we return to this below. Of central importance here is the fact that
\begin{equation}
\mathcal V_{geo} = \mathcal V_{P} \label{geop}.
\end{equation}
The geometrical and Parikh volumes are equivalent.
\subsection{Kodama Volume} \label{HaywardKodama}
In several papers (for example \cite{Hayward}, among others), Hayward defines a volume for dynamical black holes in terms of the Kodama vector, an analogue to the Killing vector in dynamical spacetimes.  He makes a similar development for cylindrical symmetry in \cite{Haywardcylinder}, but we will focus here on the spherical symmetry case as an example.

The line element for four-dimensional dynamic spherically symmetric spacetime can be written in the form
\begin{equation}
\mathrm d s^2 = g_{A B} \mathrm d x^A \mathrm d x^B + r^2 \mathrm d \Omega_2^2 \label{Haywardmetric}
\end{equation}
where there are two coordinates $x^A$ in addition to the coordinates $(\theta,\phi)$ within the 2-sphere metric $\mathrm d \Omega_2^2$.  Here, $r(x^A)$ is the aerial radius.

An analogue to the Killing vector in dynamic spherical symmetry is the Kodama vector, which we will label by $K^\alpha$.  An important property of the Kodama vector is that it becomes null on and only on the trapping horizon of a dynamic black hole, a property that is analogous to the Killing vector becoming null on and only on the Killing horizon of a stationary black hole.  The Kodama vector $K^\alpha$ is defined as the curl of the aerial radius,
\begin{equation}
K^\alpha = \epsilon^{\alpha \beta} \nabla_\beta r, \label{Kodamadef}
\end{equation}
where $\epsilon^{\alpha \beta}$ is the volume form associated with the 2-metric $g_{A B}$ from \eqref{Haywardmetric}, or
\begin{equation}
\epsilon^{\alpha \beta} = \epsilon^{A B}\delta^\alpha_A \delta^\beta_B
\end{equation}
where $\epsilon^{A B}$ is the Levi-Civita tensor for the two dimensions $x^A$ in \eqref{Haywardmetric}.  The Kodama vector agrees with the usual timelike Killing vector in stationary spherically symmetric spacetimes if $K^\alpha$ commutes with $\nabla^\alpha r$. In these cases the line element can be written as
\begin{equation}
\mathrm d s^2 = -\left( 1 - \frac{2 E(r)}{r}\right) \mathrm d t^2 + \left( 1 - \frac{2 E(r)}{r}\right)^{-1} \mathrm d r^2 +  r^2 \mathrm d \Omega_2^2 \label{Eofrmetric}
\end{equation}
where the Killing and Kodama vectors are $t^\alpha = \delta^\alpha_t$.  In general, if $E(r)$ is allowed to vary with $t$ in line elements of the form $\eqref{Eofrmetric}$ then the Kodama vector is $K = \partial_t$, though it is obviously only a Killing vector if the line-element is $t$-independent.

The Kodama vector (in spherical symmetry only) has the property
\begin{equation}
\nabla_\alpha K^\alpha = 0,
\end{equation}
which, along with the Gauss theorem (see, for example, \cite{Poisson}), implies a conserved quantity Hayward defines to be the volume,
\begin{equation}
\mathcal V_{K} = \left|\int_\Sigma K^\alpha \mathrm d \Sigma_\alpha\right| \label{HaywardV}
\end{equation}
where $\Sigma$ is a spacelike hypersurface and $\mathrm d \Sigma_\alpha$ is the volume element of the surface times a future directed normal.~(The absolute value signs are to avoid having to deal with the sign of the result.)  If the horizon of a black hole is located by $r = r_+$, then the volume can be defined as \eqref{HaywardV} with $\Sigma$ being the  region $r \leq r_+$, with result $\mathcal V_{K}=4 \pi r_+^3 / 3$. In the spherically symmetric case and with the Kodama vector the usual timelike Killing vector then
\begin{equation}
\mathcal V_{K} = \mathcal V_P = \mathcal V_{geo} \label{kc}.
\end{equation}

\subsection{Null Generator Volume} \label{OurVolumeIntro}

\begin{figure}[ht]
\epsfig{file=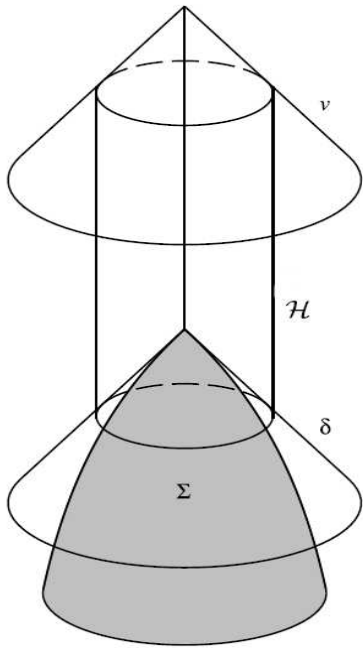,height=3.5in,width=2
in,angle=0}
\caption{The collapse of a timelike boundary surface $\Sigma$ that terminates at the central singularity simultaneously with the null cone $\delta$ and produces a black hole with horizon $\mathcal H$.  The null cone $v$ is any null cone to the future of $\delta$.  The invariant four-volume $\mathcal V$ calculated here is bounded by $\delta$ and $v$ and is to the interior of $\mathcal H$.    } \label{diagram}
\end{figure}

In \cite{arXivPaper}, we defined a volume rate for stationary non-degenerate black holes.  To define this rate, we considered a region of the black hole bounded by the event horizon and two distinct ingoing null cones, as shown schematically in Figure~\ref{diagram}.  Then, by allowing the intersection of the ingoing null cone with the event horizon to vary, it was found that the four-volume of this region grows as $\mathcal V \propto \ln \lambda$, where $\lambda$ is the affine generator of the horizon and the constant of proportionality is the volume $\mathcal{V}_{P}$ divided by the surface gravity $\kappa$ of the black hole.  Thus we define the null generator volume as
\begin{equation}\label{firstkappa}
\mathcal V_{\mathcal{N}} \equiv \frac{\mathrm d \mathcal V(\lambda)}{\mathrm d \ln \lambda}=\frac{\mathcal{V}_{P}}{\kappa}.
\end{equation}

The easiest way to demonstrate this volume is to use ingoing Eddington-Finkelstein-like coordinates. For metrics of the form (\ref{metricone}) (dropping the subscript $s$ on $t$) define $\mathrm d V \equiv \mathrm d t + \mathrm d r / \alpha(r)$, from which the line element becomes
\begin{equation}
\mathrm d s^2 = - \alpha(r) \mathrm d V^2 + 2 \mathrm d V \mathrm d r + r^2 \mathrm d \Omega_{D-2}^2.
\end{equation}
Here, $V = const.$ labels sets of ingoing null geodesics.  The $D$-volume of the region between $V = V_0$ and a larger (arbitrary) value of $V$ is
\begin{eqnarray}
\mathcal V = \int_{V_0}^V \mathrm d V' \int_0^{r_+} \mathrm d r \int \mathrm d \Omega_{D-2} \sqrt{-g}\\ \nonumber = (V - V_0) \int \mathrm d^{D-1} x \sqrt{-g},
\end{eqnarray}
and so
\begin{equation}
\frac{\mathrm d \mathcal V}{\mathrm d V} = \int \mathrm d^{D-1} x \sqrt{-g}.
\end{equation}
Assuming asymptotic flatness ($\alpha(r) \to 1$ as $r \to \infty$), the relationship between $t$ and thus $V$ and an affine parameter $\lambda$ on the event horizon is given by \cite{Poisson}
\begin{equation}
\frac{\mathrm d V}{\mathrm d \lambda} = \frac{1}{\kappa \lambda}
\end{equation}
where $\kappa$ is the surface gravity of the black hole; in this case then $V = \ln \lambda / \kappa$ up to an additive constant, and we can write
\begin{equation}
\frac{\mathrm d \mathcal V}{\mathrm d \ln \lambda} = \kappa^{-1} \int \mathrm d^{D-1} x \sqrt{-g}
\end{equation}
as claimed.

This procedure can be generalized for other non-degenerate stationary metrics, for example, in Kerr-Newman along principal ingoing null geodesics.

\section{Vector Volume} \label{vecvolintrodef}

In this section we define a vector volume rate (which we also refer to as the ``vector volume") in any space or spacetime with respect to any divergence-free vector field $v^\alpha$ ($\nabla_\alpha v^\alpha = 0$). The set of divergence-free vectors of course includes all Killing vectors. The vector volume of a $D$-dimensional region $\mathcal R$ with respect to $v^\alpha$ is written as  $\mathcal V_{v}$. The vector $v^\alpha$ must satisfy $v^{\alpha}n_{\alpha}=0$ where $n_{\alpha}$ is normal to the boundary of $\mathcal R$. We will now introduce two definitions of $\mathcal V_{v}$.  Section \ref{differentiallike} defines it as a derivative of the $D$-volume of region $\mathcal R$ along the vector field $v^\alpha$.  Section \ref{haywardlike} defines it as an integral of the vector $v^\alpha$ over a hypersurface.  These two definitions are then shown to be equivalent.  A third definition, which uses perhaps less familiar terminology, is included as Appendix \ref{vve}.

\subsection{Definition 1} \label{differentiallike}

\begin{figure}[ht]
\epsfig{file=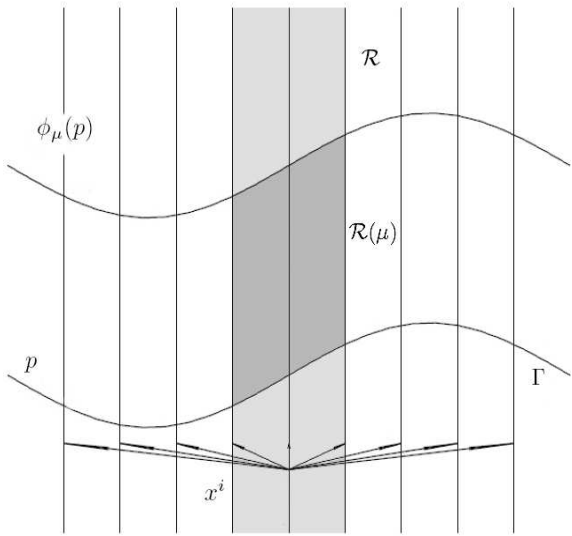,height=3in,width=3in,angle=0}
\caption{\label{diagram2} The vertical lines represent the congruence of integral curves of $v$.  The region $\mathcal R$ is shown in light grey and the specific region $\mathcal R(\mu)$ shown in darker grey.  The hypersurface $\Gamma$ is labelled and the other surface represents a movement a parameter distance $\mu$ along the integral curves. The coordinates $x^i=const$ are shown.}
\end{figure}

Define $\mathcal V(\mathcal R)$ as the $D$-dimensional volume of region $\mathcal R$. The essential point is to define a volume for which we use the derivative of the scalar volume along the vector field $v$:
\begin{equation}
v^\alpha \partial_\alpha ( \mathcal V(\mathcal R)).\label{meaningless}
\end{equation}
Unfortunately, $\mathcal V(\mathcal R)$ is not well-defined as a local quantity.  In order to provide a meaning to \eqref{meaningless} we must define $\mathcal V(\mathcal R)$ as a quantity which depends in some specific way on the coordinates. One method is to consider the congruence of integral curves of $v$.  Let $\phi_\mu(p)$ be the point lying at parameter distance $\mu$ along the integral curve of $v$ starting at point $p$.  Define $\Gamma$ as an arbitrary hypersurface of dimension $(D-1)$ which intersects every integral curve of $v$ exactly once.  Then $\Gamma \cap \mathcal R$ is a hypersurface region which lies entirely within $\mathcal R$ and intersects each integral curve within $\mathcal R$ exactly once.  Finally, define $\mathcal R(\mu)$ as the region which lies within $\mathcal R$, for which each point within $\mathcal R(\mu)$ can be expressed as
\begin{equation}
\phi_\nu(p), \;\;\;\; 0 \leq \nu \leq \mu, \;\;\;\; p \in  \Gamma \cap \mathcal R.
\end{equation}
In other words, $\mathcal R (\mu)$ is the subregion of $\mathcal R$ for which every point is at most a parameter distance of $\mu$ from $\Gamma$ along the integral curves of $v$. The situation is shown schematically in Figure \ref{diagram2}.

Since
\begin{equation}
 v^\alpha \partial_\alpha = \frac{\mathrm d}{\mathrm d \mu},
\end{equation}
from (\ref{meaningless}) we have
\begin{equation}
\mathcal V_{v} = \frac{\mathrm d}{\mathrm d \mu} \mathcal V(\mathcal R(\mu)). \label{derivativebased}
\end{equation}
Further, since $\mathcal V_{v}$ is, as we will show, a constant, this volume rate can be written as the ratio
\begin{equation}
\mathcal V_{v} = \frac{\mathcal V(\mathcal R(\mu))}{\mu}.
\end{equation}
These relations provide a simple interpretation of the vector volume: it is the growth rate of the $D$-dimensional volume $\mathcal R$ along the vector field $v$.  We show later in this section and in Section \ref{haywardlike} that $\mathcal V_{v}$ is independent of  $\mu$ and the particular choice of hypersurface $\Gamma$.

For ease of computation here we now use adapted coordinates $x^\alpha$ in which  $v^\alpha = \delta^\alpha_0$, and where $\Gamma$ is defined by $x^0 = 0$. The computation of $\mathcal V(\mathcal R(\mu))$ is now straightforward.  The region $\mathcal R(\mu)$ is simply equal to the subregion of $\mathcal R$ for which $0 \leq x^0 \leq \mu$. We can define this region as $\mathcal R(x^0)$ for $x^0 \geq 0$.  As a result, the vector volume can be written as

\begin{equation}
\mathcal V_{v} = \frac{\partial \mathcal V(\mathcal R(x^0))}{\partial x^0}.
\end{equation}

Since $\mathcal R$ is defined by the normal to its boundaries being perpendicular to $v^\alpha$, we can define $\mathcal R$ in these coordinates as $x^i \in \Sigma$ for some $(D-1)$-dimensional region $\Sigma$.  (This results in $n_\alpha$ having no $x^0$ component, confirming that $v^\alpha n_\alpha = 0$ as required.)  Then $\mathcal R(x^0)$ becomes the set of points for which $x^i \in \Sigma$ and the $x^0$ values which lie within the region vary between $0$ and the $x^0$ parameter within the expression.  $\mathcal V(\mathcal R(x^0))$ is given by the $D$-space integral,
\begin{eqnarray}
\mathcal V(\mathcal R(x^0)) = \int_0^{x^0} \mathrm d x^0 \int_\Sigma \sqrt{|g_D|} \mathrm d^{D-1} x   \\ \nonumber = x^0 \int_\Sigma \sqrt{|g_D|} \mathrm d^{D-1} x
\end{eqnarray}
where $g_D$ is the determinant of the metric $g_{\alpha \beta}$ and $\mathrm d^{D-1} x $ is the product of the differentials excepting $\mathrm d x^0$.  We note further that since $\nabla_\mu v^\mu = 0$, and using the relationship
\begin{equation}
\nabla_\mu A^\mu = |g_D|^{-\frac{1}{2}} \partial_\mu (|g_D|^{\frac{1}{2}} A^\mu)
\end{equation}
for an arbitrary vector field $A^\mu$, we find
\begin{equation}
\partial_\mu(|g_D|^{\frac{1}{2}} v^\mu) = \partial_0(|g_D|^{\frac{1}{2}}) = 0
\end{equation}
from which we find that the metric determinant is independent of coordinate $x^0$.
This implies that the vector volume is expressible as
\begin{equation}
\mathcal V_{v} = \int_{x^i \in \Sigma} \sqrt{|g_D|} \mathrm d^{D-1} x, \label{gexpress}
\end{equation}
which is independent of $x^0$ and thus $\mu$. We also note that this is the same form as the definition for the Parikh volume \eqref{Parikhdef} and of the Cveti\v{c} et al.~geometric volume \eqref{Vgeo}.  Thus the vector volume is a generalization of the Parikh and geometric volumes. This connection will be further explored in Section \ref{canon}. We can show the invariance under the choice of slicing hypersurface $\Gamma$ here but it will be easier to do so in the following section.

\subsection{Definition 2} \label{haywardlike}

The second definition is similar to that used by Hayward to construct the Kodama volume as discussed in Section \ref{HaywardKodama} and given by (\ref{HaywardV}).  Recall there that $\Sigma$ slices the region $\mathcal B$ bounded by the black hole trapping horizon (whose normal is orthogonal to the Kodama vector) and $\mathrm d \Sigma_\alpha$ is the directed surface element of $\Sigma$.  We note that the Kodama vector is, in certain situations, divergence-free.  We can generalize this definition to write the vector volume in a similar form. Replacing $K^\alpha$ with $v^\alpha$,  $\mathcal B$ with a region $\mathcal R$ whose normal is orthogonal to $v^\alpha$, and $\Sigma$ with $\Gamma \cap \mathcal R$, with $\Gamma$ as defined in the previous subsection, we have
\begin{equation}
\mathcal V_{v} = \int_{\Gamma \cap \mathcal R} v^\alpha \mathrm d \Sigma_\alpha \label{haywarddef}
\end{equation}
where here $\mathrm d \Sigma_\alpha$ is the directed surface element of $\Gamma$.  We note here that we can always choose the orientation of $\Gamma \cap \mathcal R$ such that the integral is positive.  We choose an adapted coordinate system wherein $v^\alpha = \delta^\alpha_0$, $\mathcal R$ is defined by $x^i \in \Sigma$, and $x^0 = 0$ corresponds to $\Gamma$. Then $\mathrm d \Sigma_\alpha$ is equal to $\delta^0_\alpha \sqrt{|g_D|} \mathrm d^{D-1} x$ where $\mathrm d^{D-1} x$ is the product of the differentials of $\mathrm d x^i$. The integral \eqref{haywarddef} becomes
\begin{equation}
\mathcal V_{v} = \int_{x^i \in \Sigma} \sqrt{|g_D|} \mathrm d^{D-1} x.
\end{equation}
This recovers \eqref{gexpress} and thus shows that \eqref{haywarddef} is equivalent to the definitions given in Section \ref{differentiallike}. For a parallel development using differential forms see Appendix A.

We now demonstrate that this volume is independent of the choice of particular hypersurface $\Gamma$.  To do this, we define a region $\mathcal Q$ which is bounded by two possible (nonintersecting) hypersurfaces $\Gamma$, say $\Gamma_1$ and $\Gamma_2$, as well as the boundaries of $\mathcal R$.  We now say that $\Gamma_2$ points ``outward'' and $\Gamma_1$ points ``inward,'' so that the closed integral of the vector field over the boundary of $\mathcal Q$ is given by
\begin{equation}
\oint_\mathcal{\partial Q}v^\alpha \mathrm d\Sigma_\alpha = \int_{\Gamma_2} v^\alpha \mathrm d \Sigma_\alpha - \int_{\Gamma_1} v^\alpha \mathrm d \Sigma_\alpha,
\end{equation}
where of course $v^\alpha \mathrm d \Sigma_\alpha$ is zero on $\partial \mathcal R$ since the normal is orthogonal to $v^\alpha$.  We note that the first term, by the divergence theorem, is equal to $\int_{\mathcal Q} \nabla_\alpha v^\alpha \mathrm d^{D} \mathcal V$ where $\mathrm d^{D} \mathcal V$ is the volume element $\sqrt{|g_D|} \mathrm d^{D} x$.  Rearranging, this implies
\begin{equation}
\int_{\Gamma_2} v^\alpha \mathrm d \Sigma_\alpha - \int_{\Gamma_1} v^\alpha \mathrm d \Sigma_\alpha = \int_\mathcal Q \nabla_\alpha v^\alpha \mathrm d^{D} \mathcal V.
\end{equation}
Since $\nabla_\alpha v^\alpha = 0$,
\begin{equation}\label{Gauss}
\int_{\Gamma_2} v^\alpha \mathrm d \Sigma_\alpha = \int_{\Gamma_1} v^\alpha \mathrm d \Sigma_\alpha,
\end{equation}
a standard application of Gauss' theorem.  In words, $v^{\alpha}$ is the ``flux" associated with the conservation of $\mathcal{V}_v$. However, note that for any vector field $u^\alpha$ which is not divergence free, an attempt to use the definition here for the vector volume will yield a result that is dependent on the choice of hypersurface $\Gamma$.

\section{Vector Volume - General Properties} \label{generalproperties}

Since the covariant derivative and thus the divergence are linear functions, the divergence of any linear combination of two divergence-free vectors is zero.  As a result, if $v^\alpha$ and $w^\alpha$ are valid vectors to define a vector volume, so will a linear combination of those vectors.  In what follows in this section we examine how linear combinations of choices of vector affects the resulting vector volume.

\subsection{Constant Multiplication}

Let us first show that the vector volume of a region $\mathcal R$ with respect to a non-zero positive constant $C$ times a given vector field $v$ is that same constant $C$ times the vector volume of $\mathcal R$ with respect to $v$. That is,
\begin{equation}
\mathcal V_{Cv} = C \mathcal V_{v}.
\end{equation}
To show this we use \eqref{haywarddef}.  Since the integral curves for the vector fields $v$ and $Cv$ are the same, we can use the same hypersurface region $\Gamma$ to represent the hypersurface which lies within $\mathcal R$ and intersects the integral curve(s) exactly once.  Then we can write
\begin{eqnarray}
\mathcal V_{Cv} = \int_{\Gamma\cap\mathcal R} C v^\alpha \mathrm d \Sigma_\alpha = \\ \nonumber C \int_{\Gamma\cap\mathcal R} v^\alpha \mathrm d \Sigma_\alpha = C \mathcal V_{v}
\end{eqnarray}
as required. If $C$ is a negative constant, then the result will be $\mathcal V_{Cv} = |C| \mathcal V_{v}$ because the orientation of the hypersurface, and thus the sign of $\mathrm d \Sigma_\alpha$, is always chosen so that the volume rate is positive.  We cannot choose $C = 0$ because the vector volume with respect to a zero vector field is not defined.

\subsection{Two Vectors}

Here we seek the relationship between the vector volumes of divergence-free vectors $v^\alpha$ and $w^\alpha$ and the vector volume of $v^\alpha + w^\alpha$.  We can only compare the vector volumes for $v^\alpha, w^\alpha$ and $v^\alpha + w^\alpha$ if they are volumes corresponding to a common  region $\mathcal R$.  This only occurs if the normal to $\mathcal R$ is perpendicular throughout to both $v^\alpha$ and $w^\alpha$. If we can find a case in which the boundary of $\mathcal R$ is parallel to both $v$ and $w$, and $\Gamma$ intersects the integral curves of both exactly once, while they are oriented the same direction, then the vector volume is a simple sum since
\begin{eqnarray}
\mathcal V_{v+w} = \int_{\Gamma\cap\mathcal R} (v^\alpha + w^\alpha) \mathrm d \Sigma_\alpha \\ \nonumber = \int_{\Gamma\cap\mathcal R} v^\alpha \mathrm d \Sigma_\alpha + \int_{\Gamma\cap\mathcal R} w^\alpha \mathrm d \Sigma_\alpha.
\end{eqnarray}

One possible situation in which the normal to $\partial \mathcal R$ is perpendicular to two appropriate vector fields $v^\alpha$ and $w^\alpha$ is the case in which $w^\alpha$ is a vector attached to closed, cyclic curves. Assume that we can choose an adapted coordinate system where $v^\alpha = \delta^\alpha_0$, $\eta^\alpha = \delta^\alpha_1$, and orbits of $\eta$ of length $P$ are closed, such that the points $(x^0, x^1, x^A)$ and $(x^0, x^1 + P, x^A)$ are coincident.  The most obvious instance of this is in spaces or spacetimes with azimuthal symmetry wherein $w^\alpha = \phi^\alpha$ and $P = 2\pi$ for the usual azimuthal symmetry vector $\phi^\alpha$.  A region $\mathcal R$ whose boundary normal is perpendicular to both $v^\alpha$ and $w^\alpha$ must have the form $x^A \in \Psi$ for some $(D-2)$-dimensional region $\Psi$, where there are no boundaries on $x^0$ or $x^1$.  We note that the integral curves of $v^\alpha$ can be represented by $x^i = const.$, and the integral curves of $v^\alpha + H w^\alpha$, for non-zero constant $H$, can be written as $x^A = const.$, $x^1 = H x^0 + B$ for some $B$ constant along each integral curve.  We note that the choice of $\Gamma$ as the hypersurface region $x^0 = 0$, $x^A \in \Psi$ has the property that it intersects each integral curve of $v^\alpha$ and $v^\alpha + H w^\alpha$ exactly once, so it is a suitable choice for the hypersurface $\Gamma$.  We can then calculate $\mathcal V_{v + Hw}$:
\begin{equation}
\mathcal V_{v + H w} = \int_{\Gamma\cap\mathcal R} v^\alpha \mathrm d \Sigma_\alpha + H \int_{\Gamma\cap\mathcal R} w^\alpha \mathrm d \Sigma_\alpha.
\end{equation}
In our adapted coordinates, $\mathrm d \Sigma_\alpha = \sqrt{|g_D|} \delta^0_\alpha$, so that
\begin{equation}
\int_{\Gamma\cap\mathcal R} w^\alpha \mathrm d \Sigma_\alpha = \int_{x^A \in \Psi} \sqrt{|g_D|} \delta^\alpha_1 \delta^0_\alpha = 0,
\end{equation}
and so we find that in this particular situation
\begin{equation}
\mathcal V_{v + H w} = \mathcal V_{v}. \label{twovectors}
\end{equation}
We can summarize this by saying that if $v^\alpha$ is a divergence free vector, $w^\alpha$ is a ``cyclic'' divergence-free vector (as defined above), and $\mathcal R$ is a region whose boundary normal is perpendicular to both, then the vector volume $\mathcal V_{v + H w}$ with respect to the vector $v^\alpha + H w^\alpha$ is equal to the vector volume $\mathcal V_{v}$.  Combining this result with the result of the previous subsection, we have
\begin{equation}\label{cvhw}
\mathcal V_{C v + H w} = C \mathcal V_{v}.
\end{equation}
We make use of (\ref{cvhw}) below.

\section{Kerr-Schild Metrics} \label{kerrschildtype}

A generalized Kerr-Schild spacetime (e.g.~\cite{Sopuerta}) has the form
\begin{equation}
\mathrm d s^2 = g_{\alpha \beta} \mathrm d x^\alpha \mathrm d x^\beta,
\end{equation}
where
\begin{equation}
g_{\alpha \beta} = \bar g_{\alpha \beta} + 2 K k_\alpha k_\beta, \qquad g^{\alpha \beta} = \bar g^{\alpha \beta} - 2 K k^\alpha k^\beta,
\end{equation}
where $\bar g_{\alpha \beta}$ with inverse $\bar g^{\alpha \beta}$ is some ``background'' metric, often flat space, $K$ is a scalar function, and $k_\alpha$ is a vector which is null in both the background and full metric. Further, the components of $k_\alpha$ can be raised and lowered using either metric:
\begin{equation}
g^{\alpha \beta} k_\beta = \bar g^{\alpha \beta} k_\beta = k^\alpha, \qquad g_{\alpha \beta} k^\beta = \bar g_{\alpha \beta} k^\beta = k_\alpha.
\end{equation}
Examples of spacetimes which can be expressed as Kerr-Schild metrics with a Minkowski background are all spherically symmetric spacetimes and the Kerr-Newman spacetime. Kerr-Newman-(anti) de Sitter is an example of a spacetime which can be written in Kerr-Schild form with a non-flat background (in this case, (anti) de Sitter).

The two spacetimes with metrics $g_{\alpha \beta}$ and $\bar g_{\alpha \beta}$ can be expressed in the same coordinates, and we can introduce a vector $v^\alpha$ which is well defined according to both backgrounds by setting its components to be equal.  Note that the covector associated with $v^\alpha$ is in general not the same when $v^\alpha$ is lowered by both metrics, since in general
\begin{equation}
v_\alpha \equiv g_{\alpha \beta} v^\beta = \bar g_{\alpha \beta} v^\beta + 2 K k_\alpha (k_\beta v^\beta).
\end{equation}

The Matrix Determinant Lemma (e.g.~\cite{DeterminantLemma}) states that if $A$ is an invertible square matrix and $u$ and $v$ are column vectors, then
\begin{equation}
\text{det}(A + u v^T) = (1 + v^T A^{-1} u) \text{det}(A)
\end{equation}
where $v^T$ is the transpose of vector $v$ and $A^{-1}$ is the inverse of matrix $A$. Now consider the determinant of
\begin{equation}
\bar g_{\alpha \beta} + 2 K k_\alpha k_\beta.
\end{equation}
We can represent this sum as a square matrix with the $\alpha$ index changing along the rows and the $\beta$ index changing along the columns. In this case we can represent $\bar g_{\alpha \beta}$ by $A$, $\bar g^{\alpha \beta}$ by $A^{-1}$, $2 K k_\alpha$ by $v^T$ and $k_\beta$ by $u$.  It is easy to check that $v^T A^{-1} u = \bar g^{\alpha \beta} (2 K k_\alpha k_\beta)$ in this representation.  Thus we find
\begin{eqnarray}
g = \text{det}(g_{\alpha \beta}) = \text{det}(\bar g_{\alpha \beta} + 2 K k_\alpha k_\beta) = \\ \nonumber (1 + \bar g^{\alpha \beta} 2 K k_\alpha k_\beta) \text{det}(\bar g_{\alpha \beta}) = \bar g
\end{eqnarray}
where $g$ and $\bar g$ represent the determinants of $g_{\alpha \beta}$ and $\bar g_{\alpha \beta}$ respectively.  This relies on the fact that $k_\alpha$ is null with respect to $\bar g_{\alpha \beta}$.

The equality of the determinants of the full metric and background metric implies that the volume element for the full spacetime ($\mathcal{F}$) is equal to the volume element for the background spacetime ($\mathcal{B}$):
\begin{equation}
\mathrm d^{D} \mathcal V_{\mathcal{F}} = \sqrt{|g|} \mathrm d^{D} x = \sqrt{|\bar g|} \mathrm d^{D} x = \mathrm d^{D} \mathcal V_{\mathcal{B}}.
\end{equation}
This indicates that the $D$-volume of some region can be calculated using either the full metric or the background metric.  This is not in general true for $N$-volumes where $N < D$ within the spacetime (such as lengths when $D > 1$, areas when $D>2$ etc.),~since the sub-determinants are not in general equal for $g_{\alpha \beta}$ and $\bar g_{\alpha \beta}$.

We note now that Definition 1 of the vector volume requires only a vector $v^\alpha$ and the $D$-volume of some region $\mathcal R$ parameterized by $\mu$, which is a function of the coordinates.  Since $v^\alpha$, the coordinates, and the $D$-volume are equivalent in the full spacetime and in the background spacetime, we find that the vector volume with respect to $v^\alpha$ of a region $\mathcal R$ in a Kerr-Schild spacetime $g_{\alpha \beta}$ is identical to the vector volume with respect to $v^\alpha$ of the background spacetime.  We can also see that the vector volume is the same in both the full spacetime and background using expression \eqref{gexpress} and using the equality of the two determinants.

We note before continuing that a vector which is divergence free according to $g_{\alpha \beta}$ will also be divergence-free according to $\bar g_{\alpha \beta}$:
\begin{eqnarray}
\left.\nabla_\alpha v^\alpha\right|_{\mathcal{F}} = \frac{1}{\sqrt{|g|}} \partial_\alpha (\sqrt{|g|} v^\alpha) = \\ \nonumber \frac{1}{\sqrt{|\bar g|}} \partial_\alpha (\sqrt{|\bar g|} v^\alpha) = \left.\nabla_\alpha v^\alpha\right|_{\mathcal{B}}.
\end{eqnarray}
This allows us to see that a vector which is valid for calculating the vector volume in the background spacetime will be valid for calculating the vector volume in the full spacetime.

The fact that the vector volume with respect to a full spacetime is equal to the vector volume calculated for its background spacetime is of particular importance and interest when the background spacetime is Minkowski space.  Let $\bar g_{\alpha \beta} = \eta_{\alpha \beta}$ and move for the moment into Lorentzian coordinates $(T,x^i)$ where the $x^i$ are $D-1$ spatial coordinates.  The metric can be expressed as
\begin{equation}
\eta_{\alpha \beta}\mathrm d x^\alpha \mathrm d x^\beta = - \mathrm d T^2 + \sum_i (\mathrm d x^i)^2.
\end{equation}
Now define $T^\alpha$ according to $T^\alpha \partial_\alpha \equiv \partial_T$.  Then if we have some region $\mathcal R$ defined by $x^i \in \Sigma$, where $\Sigma$ is some $(D-1)$-dimensional spatial region, the vector volume of $\mathcal R$ with respect to $T^\alpha$ in the background space is, from \eqref{gexpress},
\begin{equation}\label{euclid}
\mathcal V_{T} = \int_{x^i \in \Sigma} {|\eta|} \mathrm d^{D-1} x = \int_{x^i \in \Sigma} \mathrm d^{D-1} x \equiv \mathcal V_E,
\end{equation}
where $\mathcal V_{E}$ is the Euclidean volume of the spatial component of region $\mathcal R$ in the flat background spacetime.  Since the vector volume is equal regardless of whether we choose the flat background or the full metric, the vector volume with respect to $T^\alpha$ for the full spacetime $g_{\alpha \beta}$ will be equal to the Euclidean volume for the spatial part of the region, as calculated in the flat background.

\subsection{``Double'' Kerr-Schild Form} \label{doubleKS}

We can extend the argument given above further.  With the specific example of the Kerr-(anti) de Sitter metric, we note that it is possible to have a metric which can be expressed in Kerr-Schild form with a non-Minkowski background, which \emph{itself} can be expressed in Kerr-Schild form.  In other words, we let the full metric be expressed, as before, in the form
\begin{equation}
g_{\alpha \beta} = \bar g_{\alpha \beta} +2 K k_\alpha k_\beta, \qquad g^{\alpha \beta} = \bar g^{\alpha \beta} - 2 K k^\alpha k^\beta, \label{firstform}
\end{equation}
where, in the particular case of Kerr-(anti) de Sitter, $\bar g_{\alpha \beta}$ is the metric for (anti) de Sitter spacetime and $k_\alpha$ is a null vector with respect to both $g_{\alpha \beta}$ and $\bar g_{\alpha \beta}$.  From the previous section we have $g = \bar g$.  In the Kerr-(anti) de Sitter case, we can break this down even further by expressing the (anti) de Sitter metric in terms of the flatspace metric, say $h_{\alpha \beta}$, and a vector $l_\alpha$, which is ``null'' with respect to $h_{\alpha \beta}$ and $\bar g_{\alpha \beta}$, but not necessarily $g_{\alpha \beta}$.  The reason for this is that the (anti) de Sitter spacetime can itself be expressed in Kerr-Schild form. However, the ``null vector'' in this form will not necessarily be null in the global Kerr-(anti) de Sitter spacetime.  We can write this as
\begin{equation}
\bar g_{\alpha \beta} = h_{\alpha \beta} +2L l_\alpha l_\beta.
\end{equation}
Now we set $\bar g^{\alpha \beta}$ and $h^{\alpha \beta}$ as the inverses of $\bar g_{\alpha \beta}$ and $h_{\alpha \beta}$, and (for lack of a better name) set $\bar l^\alpha = \bar g^{\alpha \beta} l_\beta = h^{\alpha \beta} l_\beta$ as the ``contravariant'' form of $l_\alpha$ within the background spacetime (which will not, in general, be equal to $l^\alpha = g^{\alpha \beta} l_\beta$ for the full spacetime).   Note that~$\bar g = h$, where $h = \text{det}(h_{\alpha \beta})$. We know also from \eqref{firstform} that $\bar g^{\alpha \beta}$ can be written as
\begin{equation}
\bar g^{\alpha \beta} = h^{\alpha \beta} - 2L \tilde l^\alpha \tilde l^\beta, \label{contra1}
\end{equation}
and so we can decompose the spacetime into a form
\begin{equation}
g_{\alpha \beta} = h_{\alpha \beta} + 2 L l_\alpha l_\beta + 2 K k_\alpha k_\beta,
\end{equation}
where $h_{\alpha \beta}$ is the metric tensor for flat space, $2 L l_\alpha l_\beta$ is a correction from Minkowski to (anti) de Sitter, and $2 K k_\alpha k_\beta$ is a correction from (anti) de Sitter to Kerr-(anti) de Sitter.  Since $g = \bar g$ and $\bar g = h$, we find that $g = h$. The determinant of the metric tensor, and thus the associated vector volume, is the same for the Kerr-(anti)de Sitter spacetime and the background-background metric. We can now retrace the argument given above and arrive back at (\ref{euclid}).

Note that the contravariant metric equation for this ``double Kerr-Schild'' form is slightly complicated because $\tilde l^\alpha = \bar g^{\alpha \beta} l_\beta \neq l^\alpha = g^{\alpha \beta}l_\beta$; in fact $\tilde l^\alpha = l^\alpha - 2 K k^\alpha k^\beta l_\beta$.  If we set $\gamma \equiv k^\alpha l_\alpha$, we can write the contravariant metric equation somewhat compactly as
\begin{equation}
g^{\alpha \beta} = h_{\alpha \beta} -  2L (l^\alpha + 2 K k^\alpha \gamma) (l^\beta _ 2 K k^\beta \gamma) - 2 K k^\alpha k^\beta.
\end{equation}

The primary application of the vector volume is to stationary black holes, i.e.~black holes with a Killing vector corresponding to time translation which is timelike outside the black hole horizon.  We expand on two cases below. The first is the canonical black hole volume, defined and discussed in Section \ref{canon}, which is equivalent to the volume considered by Parikh as reviewed in Section \ref{Parikhintro} and the geometric volume defined by Cveti\v{c} et al.~as reviewed in Section \ref{cveticgeometricvolume} and is similar to Hayward's volume \cite{Hayward}.  The second is related to the volume introduced in Section \ref{OurVolumeIntro}. This is examined in Section \ref{ourvolume}.

\section{Canonical Black Hole Volume} \label{canon}

In this section we define and examine the ``canonical black hole volume'' for stationary spacetimes.  We define this as the vector volume of the region below the event horizon of a black hole, with respect to the canonical Killing vector which corresponds to stationarity.  We will label this volume $\mathcal V_\mathcal C$.  In general, for time coordinate $t$ we will write the corresponding Killing vector as $t^\alpha$, where $t^\alpha \partial_\alpha \equiv \partial_t$; instead of $t$ sometimes $T$ or $\tau$ or another coordinate may be used.

We note that the canonical black hole volume is equivalent to the volume that Parikh defines in his paper \cite{Parikh} as well as the geometric volume defined by Cveti\v{c} et al.~\cite{Cvetic}

To define the canonical black hole volume explicitly, assume that there is a stationary $D$-dimensional black hole which can be written in coordinates $(t,x^i)$ adapted to the Killing vector which corresponds to the black hole's stationarity, which takes the form $\partial_t$.  We also demand that $\partial_t$ be properly normalized, if possible, a condition which we elaborate on in Section~\ref{NormalizationCanon}.  Let the region below the horizon be $x^i \in \Sigma$, where $\Sigma$ is a $(D-1)$-dimensional region.  The line element can be written as
\begin{equation}
\mathrm d s^2 = g_{t t}(x^i) \mathrm d t^2 + 2 g_{t i}(x^i) \mathrm d t \mathrm d x^i + g_{i j}(x^i) \mathrm d x^i \mathrm d x^j,
\end{equation}
where, since these coordinates are adapted to the Killing vector $\partial_t$, the components of $g_{\alpha \beta}$ are independent of $t$.  If we let $g_D$ be the determinant of the metric $g_{\alpha \beta}$, then from \eqref{gexpress} we can write the canonical black hole volume as
\begin{equation}
\mathcal V_\mathcal C \equiv V_t \equiv \int_{x^i \in \Sigma} \sqrt{|g_{(D)}|} \mathrm d^{D-1} x.
\end{equation}
Compare to \eqref{Parikhdef} and \eqref{Vgeo}.

Remarkably, if $\phi^\alpha \partial_\alpha = \partial_\phi$ is a Killing vector corresponding to axial symmetry such that $\phi$ is a cyclic coordinate, it follows from \eqref{twovectors} that the vector volumes for the black hole calculated using Killing vectors $t^\alpha$ and $t^\alpha + \Omega \phi^\alpha$ are identical for any constant $\Omega$:
\begin{equation}
\mathcal V_\mathcal C = \mathcal V_t = \mathcal V_{t + \Omega \phi}. \label{canonwithomega}
\end{equation}
A consequence of this, for example, is that in stationary, axisymmetric black holes with some angular momentum, we can calculate the canonical BH volume using either the stationarity Killing vector $\partial_t$ or the Killing vector $\xi = \partial_t + \Omega_H \partial_\phi$ which is tangent to the null generators of the horizon.

An important point remains. We have to ensure that the vector $\partial_{t}$ has the proper normalization.  In his paper, Parikh suggests fixing the asymptotic normalization of the Killing vector, but this becomes problematic in spacetimes with $\Lambda$.

\subsection{Killing Vector Normalization} \label{NormalizationCanon}

Consider first an asymptotically flat stationary black hole with axial symmetry.  How should we fix the asymptotic normalization of the Killing vector corresponding to stationarity?  Following Carter \cite{LesHouches}, the normalization for the vector corresponding to stationary, say $T^\alpha$, is set by
\begin{equation}
-T^\alpha T_\alpha \to 1
\end{equation}
where the limit is taken at spatial infinity.  In spacetime with time symmetry but no spatial translational symmetry, this defines the Killing vector uniquely. (If a translational spacelike Killing vector exists, it is possible to create a new well-normalized timelike Killing vector by a linear combination of the time Killing vector with the spatial Killing vector, as in Minkowski space.)  Spacetimes with axial symmetry have a rotation Killing vector, say $m^\alpha$, which is zero on the rotation axis, whose normalization can be fixed by requiring
\begin{equation}
\frac{ \partial_\alpha X \partial^\alpha X}{4 X} \to 1
\end{equation}
where $X = m^\alpha m_\alpha$ and the limit is taken on the rotation axis.  This is what ensures the standard periodicity $2 \pi$ \cite{LesHouches}.

As explained above, any Killing vector of the form $\tilde T^\alpha \equiv T^\alpha + \Omega m^\alpha$ will yield the same canonical volume. Now the magnitude squared of $\tilde T^\alpha$ will be everywhere equal to that of $T^\alpha$ only if $\Omega = 0$, since
\begin{equation}
\tilde T^\alpha \tilde T_\alpha = T^\alpha T_\alpha + 2 \Omega T^\alpha m_\alpha + \Omega^2 X.
\end{equation}
However, since $m^\alpha = 0$ on the rotation axis, any valid Killing vector $\tilde T^\alpha$ for calculating the volume will have
\begin{equation}
-\tilde T^\alpha \tilde T_\alpha \to -T^\alpha T_\alpha \to 1
\end{equation}
with the limit taken to spatial infinity on the rotation axis.  We can use this as the condition for the normalization of the Killing vector to produce the proper canonical BH volume in axisymmetric spacetimes which are asymptotically flat.

If the space is \emph{not} asymptotically flat, then it is more challenging to define the proper normalization.  In the case of static spherical symmetry, metrics of the form (\ref{metricone}) (with $t_s=t$),
i.e.~ones wherein $g_{r r} g_{t t} = -1$, are important due to the vanishing radial null-null component of the Ricci tensor \cite{Jacobson}.  Such spacetimes include the Reissner-Nordstr\:om-(anti) de Sitter class. Now if we take $T^{\alpha}\partial_{\alpha} \equiv \partial_{t}$, the condition $g_{r r} g_{t t} = -1$ suggests that the normalization condition
\begin{equation}
-g_{r r} \tilde T^\alpha \tilde T_\alpha = 1.
\end{equation}
Here, $r$ is a uniquely defined coordinate since surfaces of $r = const.$ are spheres with surface area $4\pi r^2$ (in the 4 dimensional case).

This still leaves spacetimes which are not asymptotically flat and which do not possess spherical symmetry; an example here is Kerr-Newman-(anti) de Sitter case, which has line element
\begin{widetext}
\begin{equation}
\mathrm d s^2 = -\frac{\Delta_r}{\rho^2} \left( \mathrm d \tau - \frac{a \sin^2 \theta \mathrm d \varphi}{\Xi}\right)^2 + \frac{\Delta_\theta \sin^2\theta}{\rho^2} \left(a \mathrm d \tau - \frac{(r^2+a^2)\mathrm d \varphi}{\Xi}\right)^2 + \frac{\rho^2 \mathrm d r^2}{\Delta_r} + \frac{\rho^2 \mathrm d \theta^2}{\Delta_\theta}. \label{KdSLine}
\end{equation}
\end{widetext}
Here, $\Delta_r = \left(1 - \frac{\Lambda r^2}{3}\right)(r^2 + a^2) - 2 m r + Q^2$, $\Delta_\theta = 1 + \frac{\Lambda a^2}{3}\cos^2 \theta$, $\Xi = 1 + \frac{\Lambda a^2}{3}$, $\rho^2 = r^2 + a^2\cos^2\theta$, and $\Lambda$ is the cosmological constant, opposite in sign to that in \cite{LesHouches}. As in \cite{LesHouches} we assume here that $\Xi>0$ for $\Lambda<0$. The $r \to \infty$ limit of this metric, we note, is outside the cosmological horizon for $\Lambda>0$, and the metric does not reduce to flat space asymptotically.  Not only that, but $r$ is no longer a coordinate for which the surfaces $r = const.$ represent spheres of symmetry, but ellipsoids of symmetry instead. Nevertheless, by analogy with the requirement that a well-normalized Killing vector $\tilde T^\alpha$ has $\tilde T^\alpha \tilde T_\alpha \to -1$ on the rotation axis in asymptotically flat spacetime with axial symmetry, and that $g_{r r} g_{t t} = -1$ is a proper normalization in spherical symmetry, we are led to the following question: What if we require, in the case of Kerr-Newman-(anti) de Sitter,
\begin{equation}
-g_{r r} \tilde T^\alpha \tilde T_\alpha \to 1
\end{equation}
where the limit is taken for large $r$ on the rotation axis?  Remarkably, in the Kerr-Newman- (anti) de Sitter metric, setting $\tilde T^\alpha \partial_\alpha = \partial_\tau + \Omega \partial_\varphi$ (for any constant $\Omega$) yields
\begin{equation}
-g_{r r} \tilde T^\alpha \tilde T_\alpha = 1
\end{equation}
\emph{everywhere} on the rotation axis (where $g_{r r}$ and the norm of $\tilde T^\alpha$ are defined). This saves us the difficulty of worrying about, for example, going beyond the cosmological horizon and seems to be a reasonable normalization requirement.

There is another potential method for choosing the ``correct'' time-symmetry Killing vector for metrics which can be expressed in Kerr-Schild or ``double Kerr-Schild'' form with a flat background.

\subsection{Kerr-Schild Forms} \label{PKS}

If we have a metric which can be expressed in Kerr-Schild form or ``double Kerr-Schild form'' wherein there is a Minkowski background, another sensible way to choose the Killing vector $T^\alpha$ corresponding to stationarity is to choose a Killing vector such that, if $\eta_{\alpha \beta}$ represents the flat background,
\begin{equation}
-\eta_{\alpha \beta} T^\alpha T^\beta = 1.
\end{equation}
This definition allows us to write the background spacetime in coordinates $(T,x^i)$, where $T^\alpha\partial_\alpha = \partial_T$.  In cases where the metric is axially symmetric, we can extend this definition to any vector $\tilde T^\alpha = T^\alpha + \Omega m^\alpha$ (where $m^\alpha$ is the Killing vector for axial symmetry) by requiring
\begin{equation}
-\eta_{\alpha \beta} \tilde T^\alpha \tilde T^\beta = 1 \text{ on the rotation axis}.
\end{equation}

The advantage here is that if there is a flat background $\eta_{\alpha \beta}$, the canonical black hole volume would then, from the results of Section \ref{kerrschildtype}, be the Euclidean volume for the spatial component of the black hole, as calculated in the flat background.  As we will show in the following subsection, in the cases of static spherical symmetry in the form (\ref{metricone}), Kerr-Newman black holes, and Kerr-Newman-(anti) de Sitter black holes, this normalization scheme, and the ones presented in the previous subsection, are consistent with each other.  Thus, in these cases, we can show that the canonical black hole volume is equal to the Euclidean volume of the spatial component of the black hole region.

\subsection{Examples}

We will work in four dimensions for these examples.

\subsubsection{Spherical Symmetry}

Consider the spherically symmetric metric (\ref{metricone}). If this is a black hole, there is some value $r_+$ for which $\alpha(r_+) = 0$ and $\alpha(r) > 0$ just outside it.  Thus the black hole region is defined by $0 \leq r \leq r_+$.  We note that $t^\alpha$ with $t^\alpha \partial_\alpha = \partial_t$ is a Killing vector for which $-g_{r r} (t^\alpha t_\alpha) = 1$, so $t^\alpha$ is a valid vector for calculating the canonical black hole volume:
\begin{equation}
\mathcal V_\mathcal C = \mathcal V_{t} =  \int_{r \leq r_+} \sqrt{|g_{(D)}|} \mathrm d^{D-1} x = \frac{4 \pi r_+^3}{3},
\end{equation}
which, as Parikh noted when he did his calculation for a spherically symmetric black hole, is the Euclidean volume of a 2-sphere of radius $r_+$.

We can show that this makes sense by rewriting (\ref{metricone}) in Kerr-Schild form by choosing a new coordinate system $(T,r,\theta,\phi)$ where
\begin{equation}
\mathrm d T = \mathrm d t + \frac{1 - \alpha}{f} \mathrm d r
\end{equation}
and defining a null vector $k_\alpha$ by
\begin{equation}
k_\alpha  \mathrm d x^\alpha = \mathrm d T  + \mathrm d r.
\end{equation}
Then, setting $\eta_{\alpha \beta} \mathrm d x^\alpha \mathrm d x^\beta = - \mathrm d T^2 + \mathrm d r^2 + r^2 \mathrm d \Omega_2^2$, which is Minkowski space in spherical polar coordinates, we can write the metric as
\begin{equation}
g_{\alpha \beta} = \eta_{\alpha \beta} + (1 - \alpha(r) ) k_\alpha k_\beta,
\end{equation}
which is in Kerr-Schild form.  Further, $\eta_{\alpha \beta} t^\alpha t^\beta = -1$.  This confirms that the volume of a region with respect to $t^\alpha$ is the Euclidean volume of the spatial component of the region, as calculated in the flat background with metric tensor $\eta_{\alpha \beta}$. This provides a clear explanation for why the canonical black hole volume is the Euclidean volume for spherical black holes.

\subsubsection{Kerr-Newman}

The line element for the Kerr-Newman black hole, which is asymptotically flat, can be written in Boyer-Lindquist coordinates as
\begin{widetext}
\begin{equation}
\mathrm d s^2 = \left(\frac{\mathrm d r^2}{\Delta}+ \mathrm d \theta^2\right)\rho^2 - \left(\mathrm d t - a \sin^2 \theta \mathrm d \phi\right)^2 \frac{\Delta}{\rho^2} + \left( (r^2 + a^2) \mathrm d\phi - a \mathrm d t \right)^2 \frac{\sin^2 \theta}{\rho^2}, \label{BL}
\end{equation}
\end{widetext}
where $\rho^2 = r^2 + a^2 \cos^2\theta$, $\Delta = r^2 - 2 M r + a^2 + Q^2$ with mass $M$, charge $Q$ and rotational parameter $a$.  In these coordinates, the outer horizon lies at $r= r_+$.  The vector $t^\alpha$ with $t^\alpha \partial_\alpha = \partial_t$ is Killing and $t_\alpha t^\alpha \to -1$ asymptotically, so it is a good vector for the calculation of the canonical black hole volume.  The canonical black hole volume then is
\begin{equation}
\mathcal V_\mathcal C = \mathcal V_{t} = \int_{r \leq r_+} \sqrt{|g_{(D)}|} \mathrm d^{D-1} x = \frac{4\pi}{3} r_+ (r_+^2 + a^2).
\end{equation}

While it is not immediately obvious that the expression on the right is equal to the Euclidean volume of the background metric in Kerr-Newman, this is also the case.  The Kerr-Schild form of the Kerr-Newman black hole in Lorentzian coordinates $(T,x,y,z)$ can be written as \cite{Poisson}
\begin{equation}
g_{\alpha \beta} = \eta_{\alpha \beta} + f k_\alpha k_\beta,
\end{equation}
where $f =  \frac{r^2}{r^4 + a^2 z^2}(2 M r - Q^2)$, $\eta_{\alpha \beta} = \mathrm {diag} (-1,1,1,1)$ and $k_\alpha$ is a null vector with
\begin{equation}
k_\alpha = \left(1, \frac{r x + a y}{r^2 + a^2}, \frac{r y - a x}{r^2 + a^2}, \frac{z}{r}\right).
\end{equation}
The coordinate $r$ is the usual Boyer-Lindquist $r$ defined in these Cartesian coordinates by
\begin{equation}
\frac{x^2 + y^2}{r^2 + a^2} + \frac{z^2}{r^2} = 1.
\end{equation}
Surfaces of constant $r$ are \emph{ellipsoids} in $(x,y,z)$, where the semi-major axes are $\sqrt{r^2+a^2}$ for the $x$ and $y$ directions and $r$ for the $z$-direction.

The relationship between $T$ and the coordinates $(t,r)$ is
\begin{equation}
\mathrm d T = \mathrm d t + \frac{2 M r - Q^2}{\Delta} \mathrm d r.
\end{equation}
We can confirm that $\partial_T = \partial_t$.  Thus, we confirm that $\eta_{\alpha \beta} t^\alpha t^\beta=-1$ and so the canonical black hole volume for the Kerr-Newman black hole is equal to the the Euclidean volume of the spatial component of the flat background of the Kerr-Newman black hole when written in Kerr-Schild coordinates.

Since the horizon is at $r = r_+$, the flatspace interpretation of the volume of the region $0 \leq r \leq r_+$ is not the volume of a sphere but the volume of an \emph{ellipsoid} with semi-major axes $\sqrt{r_+^2 + a^2}, \sqrt{r_+^2 + a^2}$, and $r$ (for $x,y$ and $z$ directions respectively).  The Euclidean volume of an ellipsoid with semi-major axes $u,v$ and $w$ is $4\pi u v w/3$, which in this case is $4\pi r_+(r_+^2+a^2)/3$, confirming that the canonical black hole volume of the Kerr-Newman black hole is in fact its Euclidean volume in this set of coordinates.

\subsubsection{Kerr-Newman-(anti) de Sitter}

The line element for Kerr-Newman-(anti) de Sitter is given in \eqref{KdSLine} and in these coordinates the metric determinant $g_{(4)}$ is
\begin{equation}
g_{(4)} = - \frac{\sin^2 \theta (r^2 + a^2 \cos^2\theta)^2}{\Xi^2}.
\end{equation}
The black hole region is given by $r \leq r_0$ where $r_0$ is the appropriate root to $\Delta_r = 0$. If we use the vector $\partial_\tau$ to calculate volume as discussed in Section \ref{NormalizationCanon}, then we obtain
\begin{equation}
\mathcal V_\mathcal C = \mathcal V_{\tau} =  \int_{r \leq r_0} \sqrt{|g_{(D)}|} \mathrm d^{D-1} x = \frac{4\pi}{3} \frac{r_0 (r_0^2 + a^2)}{\Xi}
\end{equation}
Again, the expression on the right is not immediately familiar, but as explained below this corresponds to a Euclidean volume expression.

Kerr-Newman-(anti) de Sitter, with line element of the form \eqref{KdSLine}, cannot be written in ``single'' Kerr-Schild form but can be written in ``double'' Kerr-Schild form.  Introducing new coordinates $t, \phi$ defined by
\begin{eqnarray}
\mathrm d t = \mathrm d \tau + \frac{2 m r - Q^2} {(1-\lambda r^2) \Delta_r }\mathrm d r, \\ \nonumber \qquad \mathrm d \phi = \mathrm d \varphi - \lambda a \mathrm d \tau + \frac{ (2 m r - Q^2) a r }{(r^2+a^2) \Delta_r} \mathrm d r,
\end{eqnarray}
the line element for the full spacetime overall can be written in terms of the background (anti) de Sitter metric by writing
\begin{equation}
\mathrm d s^2 = \mathrm d \bar s^2 + \frac{2 m r - Q^2}{\rho^2} (k_\alpha \mathrm d x^\alpha)^2,
\end{equation}
where
\begin{equation}
k_\alpha \mathrm d x^\alpha = \frac{\Delta_\theta \mathrm d t}{\Xi} + \frac{\rho^2 \mathrm d r}{(1 - \lambda r^2)(r^2+a^2)} - \frac{a \sin^2 \theta \mathrm d \phi}{\Xi},
\end{equation}
\begin{widetext}
\begin{equation}
\mathrm d \bar s^2 = - \frac{\Delta_\theta ( 1 - \lambda r^2)}{\Xi} \mathrm d t^2 + \frac{\rho^2}{(1 - \lambda r^2)(r^2+a^2)} \mathrm d r^2 + \frac{\rho^2}{\Delta_\theta} \mathrm d \theta^2 + \frac{r^2+a^2}{\Xi} \sin^2 \theta \mathrm d \phi^2, \label{dbars}
\end{equation}
\end{widetext}
$\lambda = \Lambda/3$ and $\rho^2, \Delta_\theta, r$ and $a$ have the same meanings as in \eqref{KdSLine} \cite{GibbonsKdS}.  It is not readily apparent that the expression for $\mathrm d \bar s^2$ is the line element for (anti) de Sitter spacetime, but by making the substitutions defined implicitly by \cite{AckayMatzner}
\begin{equation}
R^2 = \frac{r^2 \Delta_\theta + a^2 \sin^2\theta}{\Xi}, \qquad R \cos \Theta = r \cos\theta,
\end{equation}
we find that $\mathrm d \bar s^2$ becomes
\begin{equation}
\mathrm d \bar s^2 = -(1 - \lambda R^2) \mathrm d t^2 + \frac{\mathrm d R^2}{1 - \lambda R^2} + R^2 (\mathrm d \Theta^2 + \sin^2\Theta \mathrm d \phi)
\end{equation}
which is the familiar form with radial coordinate $R$ and angular coordinate $\Theta$.  By defining coordinate $T$ by
\begin{equation}
\mathrm d T  = \mathrm d t + \frac{\lambda R^2}{1 - \lambda R^2} \mathrm d R
\end{equation}
we can rewrite $\mathrm d \bar s^2$ as
\begin{equation}
\mathrm d \bar s^2 = \mathrm d s^2_{flat} + \lambda R^2 (l_\alpha \mathrm d x^\alpha)^2
\end{equation}
where $l_\alpha \mathrm d x^\alpha = \mathrm d T + \mathrm d R$ (which is a null vector in the (anti) de Sitter background and the Minkowski background) and $\mathrm d s^2_{flat} = -\mathrm d T^2 + \mathrm d R^2 + R^2 (\mathrm d \Theta^2 + \sin^2\Theta \mathrm d \phi^2)$ which is the familiar form for Minkowski space in spherical polar coordinates.  This means that the full Kerr-Newman-(anti) de Sitter metric becomes
\begin{equation}
g_{\alpha \beta} = \eta_{\alpha \beta} + \lambda R^2 l_\alpha l_\beta + \frac{2 m r - Q^2}{\rho^2} k_\alpha k_\beta
\end{equation}
where $\eta_{\alpha \beta}$ is the Minkowski metric, $k_\alpha$ is a null vector in the full Kerr-Newman-(anti) de Sitter spacetime and the (anti) de Sitter background, and $l_\alpha$ is a null vector in the (anti) de Sitter background and the flat background for the (anti) de Sitter background.

If we call the Killing vector used to define the canonical black hole volume $t^\alpha \partial_\alpha = \partial_\tau$ from before, we note that it is also equal to $t^\alpha \partial_\alpha = \partial_\tau = \partial_t = \partial_T$.  Thus we note immediately that $-\eta_{\alpha \beta} t^\alpha t^\beta = 1$ everywhere, which is the suggestion made in Section \ref{PKS} for dealing with spacetimes such as Kerr-(anti) de Sitter for which asymptotic properties are not well defined.  According to the results of Section \ref{doubleKS}, as well, the vector volume of a region in a ``double'' Kerr-Schild metric for which we use $\partial_T$ as the vector will yield the Euclidean spatial volume of that region in the flat background.  As a result, if we use $\partial_t = \partial_T$ for the vector to calculate the canonical black hole volume we will again calculate the Euclidean volume of the flat background.

In the case of Kerr-Newman-(anti) de Sitter, the black hole region is bounded by one of the solutions to $\Delta_r  = 0$; let us call the solution which gives the event horizon $r = r_0$.  The surface defined by $r = r_0$ is not a sphere in the Minkowski background, because the Minkowski background has $R$, not $r$, as a radial coordinate.  If we define Euclidean coordinates ($x,y,z$) by $x = R \sin \Theta \cos \phi, y = R \sin \Theta \sin\phi, z = R \cos \Theta$ in the usual way, we find a relationship between $(x,y,z)$ and $r$ as
\begin{equation}
\frac{x^2 + y^2}{\Xi^{-1} (r^2+a^2)} + \frac{z^2}{r^2} = 1.
\end{equation}
This implies that surfaces of constant $r$ are ellipsoids with two semi-major axes $\sqrt{ (r^2+a^2)/\Xi}$ and one semi-major axis $r$.  The Euclidean volume inside the region bounded by $r = r_0$ then would be $4\pi r_0 (r_0^2+a^2)/3\Xi$ as we discovered.  Once again the canonical black hole volume corresponds to the Euclidean volume of the black hole region.

\subsection{Connection to Area}

The invariant surface area of the horizon $\mathcal{A_H}$ (calculated in the usual way) of a spherically symmetric black hole with horizon radius $r_+$ is simply $4\pi r_+^2$.  The surface area of a Kerr-Newman black hole horizon is $4 \pi (r_+^2 +a^2)$ and the surface area of a Kerr-Newman-(anti) de Sitter black hole horizon is $4 \pi r_+ (r_+^2 + a^2) / \Xi$.  We note then that in all three cases the ratio of the canonical black hole volume $\mathcal V_C$ to the black hole surface area is:
\begin{equation}
\mathcal V_\mathcal C = \mathcal {V}_{geo} =\frac{r_+ \mathcal A_H}{3}, \label{volarearatio}
\end{equation}
which recovers \eqref{VrA1}.

In spherical symmetry with coordinates of a form like (\ref{metricone}), surfaces of constant $t, r$ have surface area $4 \pi r^2$ for all $r$.  However, in the Kerr-Newman (and Kerr-Newman-(anti) de Sitter) case the area only has the above form on the horizons; for example, in Kerr-Newman, taking a 2-surface $\Gamma$ defined by $r= R, t = const.$ gives a line element (from \eqref{BL})
\begin{eqnarray}
\mathrm d s_\Gamma^2 = (R^2 + a^2 \cos^2\theta) \mathrm d \theta^2 + \\ \nonumber \frac{\left((R^2+a^2)^2 - \Delta(R) a^2 \sin^2\theta\right)\sin^2\theta}{R^2 + a^2 \cos^2\theta}\mathrm d \phi^2,
\end{eqnarray}
from which the area is
\begin{equation}
\mathcal{A} = \int \sqrt{(R^2+a^2)^2 - \Delta(R) a^2 \sin^2\theta} \sin\theta \mathrm d \theta \mathrm d \phi.
\end{equation}
For simplicity, define
\begin{equation}
B \equiv \frac{\Delta a^2}{(R^2 + a^2)^2},
\end{equation}
which will be zero only on the horizons.  Then the area of a surface of constant $r,t$ can be written as
\begin{widetext}
\begin{equation*}
\mathcal{A} = \left\{
\begin{array}{rl}
2\pi(R^2 + a^2) \left( 1 + \left( |B|^{-\frac{1}{2}} + |B|^{\frac{1}{2}}\right)\arcsin\sqrt{\frac{B}{B-1}}\right) & \text{if } B < 0\\
\\
4\pi(R^2 + a^2) & \text{if } B = 0\\
\\
2\pi (R^2 + a^2) \left( 1 + \left( B^{-\frac 1 2} - B^{\frac 1 2}\right)\mathrm{arcsinh}\sqrt{\frac{B}{1-B}} \right) & \text{if } 0 < B \leq 1 \\
\\
2\pi(R^2 + a^2)\left( 1 + \left(B^{\frac{1}{2}} - B^{-\frac{1}{2}}\right)\left( \frac \pi 2 - \ln\left(1 - \frac{1}{\sqrt{B}} \right)\right)\right) & \text{if } B > 1
\end{array} \right.
\end{equation*}
\end{widetext}
This shows that the relationship \eqref{volarearatio} is a property of the horizon alone, since $B=0$ if and only if $\Delta = 0$.

We note also that the area is calculated through very different means than the vector volume.  While the canonical black hole volume, for example, turns out to be equal to the Euclidean volume of the flat space background, the surface area of the horizon in Kerr-Newman and Kerr-Newman-(anti) de Sitter is not what is expected from the surface area of ellipsoids in Euclidean space.  In Kerr-Newman, wherein the Minkowski representation of $r = R, t = const.$ is an ellipsoid with one semi-axis $R$ and two semi-axes $\sqrt{R^2 + a^2}$, the Euclidean surface area of these 2-surfaces is
\begin{equation}
\mathcal{A}_{E} = \pi \left(2( R^2 + a^2) + \frac{R^2}{e}\ln\left(\frac{1 + e}{1 - e}\right)\right)
\end{equation}
where $e=a/\sqrt{R^2+a^2}$ is the ellipticity of the ellipsoid \cite{Mathworld}.

\section{Null Generator Volume and Surface Gravity} \label{ourvolume}

As stated in Section \ref{OurVolumeIntro}, in \cite{arXivPaper} we defined a black hole volume by
\begin{equation}
\mathcal V_{\mathcal{N}}= \frac{\mathrm d \mathcal V}{\mathrm d \ln \lambda} \nonumber
\end{equation}
where $\lambda$ is the null affine generator on the horizon.  We will now show that in stationary, axially symmetric black holes this volume is equal to the vector volume of the black hole region with respect to the (unique) Killing vector $k^\alpha$ for which
\begin{equation}
k^\beta \nabla_\beta k^\alpha = k^\alpha
\end{equation}
on the horizon.

Spacetimes which are stationary and axially symmetric permit a Killing vector $\xi^\alpha$ which is tangent to the null generators of the horizon, for which (again, on the horizon only) \cite{Poisson}
\begin{equation}
\xi^\beta \nabla_\beta \xi^\alpha = \kappa \xi^\alpha \label{horizonequation}
\end{equation}
where $\kappa$ is the surface gravity of the horizon.  Generally speaking, $\xi^\alpha$ is written as
\begin{equation}
\xi^\alpha = t^\alpha + \Omega_H \phi^\alpha
\end{equation}
where $t^\alpha$ is the Killing vector corresponding to stationarity and $\phi^\alpha$ the Killing vector corresponding to axial symmetry, both of which have the properties explained in Section \ref{NormalizationCanon}, and $\Omega_H$ is the angular velocity which is a constant.

If $\xi^\alpha$ is multiplied by a non-zero constant $K$, then \eqref{horizonequation} becomes
\begin{equation}
 K\xi^\beta \nabla_\beta (K \xi^\alpha) = K^2 \kappa \xi^\alpha = K \kappa (K \xi^\alpha)
\end{equation}
which means that for the new vector $K \xi^\alpha$, we effectively have a new value for the constant of proportionality which was once held by $\kappa$.  If we define $K = 1/\kappa$ and set
\begin{equation}
k^\alpha \equiv \xi^\alpha/\kappa = (t^\alpha + \Omega_H \phi^\alpha)/\kappa, \label{kdefinition}
\end{equation}
then we have
\begin{equation}
k^\beta \nabla_\beta k^\alpha = k^\alpha. \label{unitycondition}
\end{equation}
This is the unique Killing vector which is both proportional to the null generators on the horizon and which satisfies \eqref{unitycondition} on the horizon.  (If the horizon is degenerate this vector will not exist since the Killing vectors tangent to the horizon have $\nabla_\beta \xi^\alpha = 0$ and $\kappa = 0$ in this case.)

Since $k^\alpha$ is a Killing vector, we can choose a system of adapted coordinates $(\bar k, x^i)$ where $k^\alpha \partial_\alpha = \partial_{\bar k}$.  Since $k^\alpha$ is tangent to the null generators on the horizon, on the horizon we have $x^i = const., \bar k = \bar k(\lambda)$, where $\lambda$ is an affine parameter.  The affine parameter can be found from \cite{Poisson}
\begin{equation}
\frac{\mathrm d \lambda}{\mathrm d \bar k} = \exp \left[\int^{\bar k } \mathrm d {\bar k'} \right] = \exp \bar k,
\end{equation}
which, rearranging, implies that up to a linear transformation,
\begin{equation}
\bar k = \ln \lambda.
\end{equation}

Writing $k^\alpha \partial_\alpha = \mathrm d / \mathrm d \ln \lambda$, from \eqref{derivativebased} we have
\begin{equation}
\mathcal V_{k} = \frac{\mathrm d \mathcal V_{\mathcal B}}{\mathrm d \ln \lambda}
\end{equation}
where $\mathcal V_{k}$ is the vector volume of the black hole region $\mathcal B$ with respect to the vector $k^\alpha$, and $\lambda$ is the affine parameter on the horizon.
The advantage of using this as the volume of stationary black holes is that the choice of vector $k$ relies only on local parameters, rather than, as with the canonical black hole volume, requiring asymptotic flatness or a Kerr-Schild form.

The relationship between the volume generated by $k^\alpha$ and the canonical black hole volume can be found by using \eqref{kdefinition}.
With the help of (\ref{cvhw}) we find that for a region $\mathcal{B}$
\begin{equation}
\mathcal V_{k} = \mathcal V_{\kappa^{-1} t + \kappa^{-1}\phi} = \kappa^{-1} \mathcal V_{t},
\end{equation}
in agreement with (\ref{firstkappa}).
This allows us to give a new definition for the surface gravity $\kappa$ in terms of the ratio of these volumes:
\begin{equation}
\kappa \equiv \frac{\mathcal V_{t}}{\mathcal V_{k}} = \frac{\mathcal V_\mathcal C}{\mathcal V_\mathcal N}, \label{newkappadefinition}
\end{equation}
where the second equality makes explicit that the surface gravity is the ratio of the canonical and null-generator volumes.  In the special case where the spacetime is a Kerr-Schild spacetime with a flat background and $\eta_{\alpha \beta} t^\alpha t^\beta = -1$, then $\mathcal V_{t} \equiv \mathcal V_{E}$, the Euclidean spatial volume, calculated in the background space. We then have
\begin{equation}
\kappa \equiv \frac{\mathcal V_{E}}{\mathcal V_{k}}.
\end{equation}

The usual physical meaning given to the surface gravity is, as explained in \cite{Poisson}, ``the force required of an observer at infinity to hold a particle (of unit mass) stationary at the horizon.'' The interpretation given here, that the surface gravity is the ratio of the canonical black hole volume (and for Kerr-Schild spacetimes, the Euclidean spatial volume of the background spacetime) to the rate of change with respect to the logarithm of the affine parameter of the invariant four-volume of a black hole, is a local interpretation that would appear to be new.


An important question is, can we use this to gain further insights into black hole mechanics? The most obvious conclusion regards the third law of black hole mechanics, $\kappa \nrightarrow 0$. Since $\mathcal V_{t} > 0$ even in the degenerate case (consider, for example, the static spherically symmetric case) we see that the third law demands that the rate of growth $\mathcal V_{k}$ must remain finite.  In order to violate the third law we need $\mathcal V_{k} \to \infty$ and since $\mathrm d \mathcal V_{\mathcal{B}}/\mathrm d \lambda$ is finite, this requires $\lambda \to \infty$ in agreement with the formulation of Israel \cite{Israel}. That is, in a sequence of quasi-static steps, the reduction of $\kappa$ to zero would take infinite advanced time.

\section{Connection to Other Authors} \label{otherauthors}

Here we further discuss the connection between the vector volume and some of the other works referenced in Section \ref{review}.

\subsection{Cveti\v{c} et al.}

Following the geometrical approach introduced in \cite{KastorEtal:2009}, and commenting here specifically on the detailed discussion in \cite{Cvetic}, it should be clear now that the geometric volume defined in \cite{Cvetic} as quoted in \eqref{Vgeo} is equal to the canonical black hole volume and thus is an instance of the use of the more general vector volume.  Additionally, they noted the relationship \eqref{VrA1} which is a higher-dimensional generalization of \eqref{volarearatio}. Of particular interest is the modified Smarr-Gibbs-Duhem relation (\ref{sgd}) and the thermodynamic volume relation (\ref{vthermo}). These results involve integrals over an infinite $(D-2)$-surface with modifications to remove the contribution from $\Lambda$ to the $E$ and $J$ integrals.  In particular, $\mathcal{V}_{th}$ is eventually derived as an integral over a 2-surface of the Killing potential for $\chi^\alpha$, the Killing vector proportional to the null generators on the horizon.

In order to demonstrate how the vector volume enters into (\ref{sgd}) in a somewhat natural way, we demonstrate a somewhat similar relation, modified in such a way as to focus on the vector volume.  The derivation here largely follows \cite{Poisson}.

For a hypersurface $\Sigma$ with boundary $S$, one form of Gauss' theorem relating to an antisymmetric tensor $B^{\alpha \beta}$ is
\begin{equation}
\int_\Sigma \nabla_\beta B^{\alpha \beta} \mathrm d \Sigma_\alpha = \frac{1}{2}\oint_S B^{\alpha \beta} \mathrm d S_{\alpha \beta}
\end{equation}
where $\mathrm d \Sigma_\alpha$ is the volume element and $\mathrm d S_{\alpha \beta}$ is the surface element.  For a Killing vector $\xi^\alpha$, $\nabla_\beta \nabla^\beta \xi^\alpha = -R^\alpha_\beta \xi^\beta$.  Additionally, $\nabla^\beta \xi^\alpha$ is an antisymmetric tensor.  As a result, for Killing vectors we can write Gauss' law as
\begin{equation}
\oint_S \nabla^\alpha \xi^\beta \mathrm d S_{\alpha \beta} = 2 \int_\Sigma R^\alpha_\beta \xi^\beta \mathrm d \Sigma_\alpha \label{Rtensor}
\end{equation}

Now let the hypersurface $\Sigma$ be a particular hypersurface spanning the black hole region. Its outer boundary is the horizon $H$.  Let its inner boundary be defined by $S'$, which we let be an arbitrarily small surface which encloses the singularity; in the Kerr-Newman-(anti) de Sitter class of spacetimes, we can define this by $t = const.,$ $r = \delta$ where we let $\delta \to 0$.  Now we use \eqref{Rtensor} in this case with our vector $k$ as defined in Section \ref{ourvolume}, noting that the integral over the boundary surface will consist of two parts---one for the horizon $H$ and one for the inner surface $S'$.
\begin{equation}
\int_\Sigma R^\alpha_\beta k^\beta \mathrm d \Sigma_\alpha = \frac{1}{2} \left(\oint_H \nabla^\alpha k^\beta \mathrm d S_{\alpha \beta} - \oint_{S'} \nabla^\alpha k^\beta \mathrm d S_{\alpha \beta}\right) \label{areavolume}
\end{equation}
For ease of representation, let $I_H = \oint_H \nabla^\alpha k^\beta \mathrm d S_{\alpha \beta}$ and $I_{S'} = \oint_{S'} \nabla^\alpha k^\beta \mathrm d S_{\alpha \beta}$.

We can now use Einstein's equations,
\begin{equation}
R^\alpha_\beta = 8\pi \left(T^\alpha_\beta - \frac{1}{2}T \delta^\alpha_\beta\right) + \Lambda \delta^\alpha_\beta
\end{equation}
to rewrite the left hand side of (\ref{areavolume}) as
\begin{equation}
8\pi \int_\Sigma T^\alpha_\beta k^\beta \mathrm d \Sigma_\alpha - 4\pi \int_\Sigma T k^\alpha \mathrm d \Sigma_\alpha + \Lambda \int_\Sigma k^\alpha \mathrm d \Sigma_\alpha,
\end{equation}
where of course $\int_\Sigma k^\alpha \mathrm d \Sigma_\alpha = \mathcal V_{k}$, volume of $\Sigma$, which, as $S'$ becomes smaller, approaches the volume of the black hole.  Let us define the integral over the stress-energy tensor as $I_T$,
\begin{equation}
I_T \equiv 4\pi \int_\Sigma \left(2T^\alpha_\beta k^\beta - T k^\alpha\right)\mathrm d \Sigma_\alpha.
\end{equation}
The left-hand side of \eqref{areavolume} is thus equal to $\Lambda \mathcal V_{k} + I_T$.  We note that in vacuum (or vacuum with $\Lambda$), $I_T$ will be identically zero.

We now find $I_H$.  As discussed in Section \ref{ourvolume}, on $H$ the Killing vector $k^\alpha$ has the property that it is tangent to the null generators, $k^\beta \nabla_\beta k^\alpha = k^\alpha$.  We can write $\mathrm d S_{\alpha \beta} = 2 k_{[\alpha} N_{\beta]} \mathrm d S$, where the square brackets denote anti-symmetrization and $N_\beta$ is an auxiliary null vector defined by $N^\alpha k_\alpha = -1$ and where $N^\alpha$ is orthogonal to the vectors $e^\alpha_A = \partial x^\alpha / \partial \theta^A$, where the $\theta^A$ represent coordinates on the horizon.  In this case, following a procedure similar to that given in \cite{Poisson}, we can define the area $\mathcal A$ in terms of $I_H$ as follows:
\begin{widetext}
\begin{equation}
I_H = \oint_H \nabla^\alpha k^\beta (2 k_{[\alpha} N_{\beta]}) \mathrm d S = 2 \oint_H k^\alpha \nabla_\alpha k^\beta N_\beta \mathrm d S = 2 \oint_H k^\beta N_\beta \mathrm d S = -2\mathcal A,
\end{equation}
\end{widetext}
where in the last equality we used $k^\beta N_\beta = -1$ and that the integral of the surface element over the surface is $\mathcal A$.

To define $I_{S'}$, we first review the definitions of mass and angular momentum from the Komar formulae.  The black hole mass $M_H$ and angular momentum $J_H$ can be defined using Komar formulae as integrals over the horizon:
\begin{equation}
M_H = - \frac{1}{8\pi} \oint_H \nabla^\alpha t^\beta \mathrm d S_{\alpha \beta} \label{komarm}
\end{equation}
and
\begin{equation}
J_H = \frac{1}{16\pi} \oint_H \nabla^\alpha \phi^\beta \mathrm d S_{\alpha \beta}, \label{komarj}
\end{equation}
where $t^\alpha$ is the well-normalized time Killing vector and $\phi^\alpha$ is the well-normalized axial symmetry Killing vector.
These appear in the Smarr formula for stationary black holes along with the surface gravity $\kappa$ and the surface area of the horizon $A$,
\begin{equation}
M_H - 2 \Omega_H J_H = \frac{\kappa \mathcal A}{4\pi}.
\end{equation}
Along similar lines, then, define $M_{S'}$ and $J_{S'}$ by integrating over the limiting surface around the horizon, $S'$, instead of over the horizon:
\begin{equation}
M_{S'} = - \frac{1}{8\pi} \oint_{S'} \nabla^\alpha t^\beta \mathrm d S_{\alpha \beta} \label{komarms}
\end{equation}
\begin{equation}
J_{S'} = \frac{1}{16\pi} \oint_{S'} \nabla^\alpha \phi^\beta \mathrm d S_{\alpha \beta}. \label{komarj1}
\end{equation}
Since $k^\alpha = \kappa^{-1} (t^\alpha + \Omega_H \phi^\alpha)$, where $\kappa$ is the surface gravity and $\Omega_H$ the angular velocity of the black hole, we can now write
\begin{equation}
I_{S'} = \oint_{S'} \nabla^\alpha k^\beta \mathrm d S_{\alpha \beta} = \frac{-8 \pi M_{S'} + 16 \pi \Omega_H J_{S'}}{\kappa}.
\end{equation}

As a result, \eqref{areavolume} gives rise to a modified Smarr relation,
\begin{equation}
\Lambda \mathcal V_{k} + I_T = -\mathcal A +\frac{4\pi M_{S'}}{\kappa} - \frac{8 \pi \Omega_H J_{S'}}{\kappa}.
\end{equation}
If we now rewrite the volume term in terms of the volume $\mathcal V_{t} = \kappa \mathcal V_{k}$, this expression becomes
\begin{equation}
M_{S'} = \frac{\kappa \mathcal A}{4\pi} + 2 \Omega_H J_{S'} + \frac{\Lambda \mathcal V_{t}}{4 \pi} + \frac{\kappa I_T}{4 \pi}. \label{ourSmarr}
\end{equation}

To confirm that these definitions of mass and angular momentum might have meaning, we first check them in four-dimensional Schwarzschild-(anti) de Sitter space
($D=4$ and $\alpha(r) = 1 - 2 m /r - \Lambda r^2/3$ in (\ref{metricone})). We find that $M_{S'}$ approaches $m$ (as the $r = const.$ surface approaches $r = 0$), whereas $M_H = m - \Lambda r_+^2/3$, where $r_+$ is the value of $r$ on the horizon.  This helps give some weight to the definitions as presented.  In Kerr spacetime (line element \eqref{BL} with $Q = 0$), $M_H = M_{S'} = m$ and $J_H = J_{S'} = m a$.  In Kerr-(anti)de Sitter (line element (\ref{KdSLine}) with $Q=0$), we find a slightly different form,
\begin{equation}
M_{S'} = \frac{m\left(1 - \Lambda a^2 / 3\right)}{\Xi^2}, \qquad J_{S'} = \frac{m a}{\Xi^2}.
\end{equation}
Equation (\ref{ourSmarr}) is in fact very similar in form to (\ref{sgd}). In cases where the charge is non-zero, the Komar formulae integrated over $S'$ diverge and so some method to subtract out the charge contribution would need to be introduced.

\subsection{Kodama Vector}
Since the Kodama vector from Section \ref{HaywardKodama} has a zero expansion in spherical symmetry, Hayward's Kodama volume \eqref{HaywardV},
wherein $K^\alpha$ is orthogonal to the boundaries of $\Sigma$, is clearly, in the case of spherical symmetry, a vector volume.  If we write the metric for dynamic spherical symmetry in the form \eqref{Eofrmetric}, then $g_{r r} g_{t t} = -1$, which explains in some measure why the Kodama volume for the region $0 \leq r \leq r_0$ is equal to the Euclidean volume of a sphere of $r=r_0$.  This implies that the Kodama volume is potentially a sensible generalization of the canonical black hole volume.

\section{Conclusion} \label{conclusion}

We have defined a vector volume in spacetime and have shown that this volume is a conserved, invariant quantity with several notable properties.  We defined a canonical black hole volume and showed that in Kerr-Schild metrics with a Minkowski background,  this volume corresponds to the Euclidean volume of the spatial component of the black hole region.  We have shown that the work of Parikh, Cveti\v{c} et al.~and Hayward involve the use of specific instances of the vector volume.  In addition to these, we proposed a null generator volume for non-degenerate stationary black holes. This volume has the advantage that it depends on neither the asymptotic properties of the spacetime nor the background metric in Kerr-Schild type spacetimes. Combining the canonical black hole volume and null generator volume, both of which are special cases of the vector volume, we have arrived at a new definition of the surface gravity.

\begin{acknowledgments}
We thank Maulik Parikh who pointed out \cite{Parikh} in response to our early draft in \cite{arXivPaper}.  This work was supported in part a grant (to KL) from the Natural Sciences and Engineering Research Council of Canada. Portions of this work were made possible by the use of \emph{GRTensorII} \cite{GrTensor}.
\end{acknowledgments}

\appendix

\section{Vector volume element via differential forms} \label{vve}

This is a slightly more abstract version of the definition given in Section \ref{haywardlike}. We refer the reader to the readable account on differential forms by Israel \cite{Israel1}. The volume \eqref{haywarddef} can be expressed in terms of differential forms as follows.  We let $\epsilon$ be the Levi-Civita tensor, or volume $D$-form, such that the $D$-volume $\mathcal V$ of a generic region $\mathcal Q$ is
\begin{equation}
\mathcal V_\mathcal Q = \int_\mathcal Q \epsilon.
\end{equation}
We now define a  $(D-1)$-form ``vector volume element,'' $\delta \mathcal V_v$, by
\begin{equation}
\delta \mathcal V_v \equiv i_v \epsilon
\end{equation}
where $i_X \alpha$ is the $(n-1)$-form interior product of the $n$-form $\alpha$ with the vector $X$.  We then define the vector volume by integrating this $(D-1)$-form along the hypersurface region $\Gamma$ as defined in Section \ref{differentiallike}:

\begin{equation}
\mathcal V_{v} = \int_{\Gamma\cap\mathcal R} \delta \mathcal V_v = \int_{\Gamma\cap\mathcal R} i_v \epsilon.
\end{equation}

In adapted coordinates where $v^\alpha = \delta^\alpha_0$, $\epsilon_{\alpha \beta \gamma \ldots \mu} = \sqrt{|g_D|} [\alpha \: \beta \: \gamma \ldots \mu]$, where $ [\alpha \: \beta \: \gamma \ldots \mu]$ is equal to $1$ ($-1$) if $\alpha,\beta,\gamma\ldots \mu$ is an even (odd) permutation of $0,1,2\ldots D-1$ and zero otherwise.  This implies that $i_v \epsilon$ is a $(D-1)$-form with components
\begin{eqnarray}
(i_v \epsilon)_{\alpha \beta \ldots \mu} = v^\nu \epsilon_{\nu \alpha \beta \ldots \mu} = \epsilon_{0 \alpha \beta \ldots \mu} = \\ \nonumber \sqrt{|g_D|} [0 \: \alpha \: \beta \ldots \mu].
\end{eqnarray}

The definition of the integral of an $n$-form over an $n$-dimensional manifold is given by Wald \cite{Wald} as follows.  If there exist coordinates $x^0, x^1, \ldots x^{n-1}$ on the manifold, then an $n$-form $\alpha$ defined on the manifold can be written in the form $\alpha = a \mathrm d x^0 \wedge \mathrm d x^1 \wedge \ldots \mathrm d x^{n-1}$, where the wedge symbol denotes a totally antisymmetric product and $a$ is a scalar.  The integral of $\alpha$ is then defined as
\begin{equation}
\int \alpha = \int a \mathrm d x^0 \mathrm d x^1 \ldots \mathrm d x^{n-1},
\end{equation}
or the integral of the scalar over the product of the differentials.

The manifold over which $i_v \epsilon$ is defined is the hypersurface $\Gamma$, and we can use coordinates $x^1, x^2, \ldots x^{D-1}$ for the $(D-1)$-dimensional hypersurface.  In these coordinates, we can write
\begin{equation}
i_v \epsilon = \sqrt{|g_D|} \: \mathrm d x^1 \wedge \mathrm d x^2 \ldots \mathrm d x^{D-1},
\end{equation}
so that
\begin{equation}
\int_{\Gamma\cap\mathcal R} \mathcal V_v = \int_{\Gamma\cap\mathcal R} i_v \epsilon = \int_{x^i \in \Sigma} \sqrt{g_D|} \mathrm d^{D-1} x,
\end{equation}
and we recover \eqref{gexpress}.  $\mathcal V_v$, as the interior product of the vector field $v$ with the volume form $\epsilon$, has a very clear interpretation, which is the advantage of presentation given in this Appendix.

We can now show that the reason for the invariance of $\mathcal V_{v}$, under the choice of $\Gamma$, is because the contribution of the ``vector volume element'' from each individual integral curve of $v$ is the same regardless of its position along the curve.  To demonstrate this, we take the Lie derivative $\mathcal L_v$ of the vector volume element along the vector field $v$, in its formulation as a $(D-1)$-form $i_v \epsilon$. This gives
\begin{equation}
\mathcal L_v (\delta \mathcal V_v) = \mathcal L_v (i_v \epsilon) = \text{div}(v) \text{ } i_v \epsilon, \label{liev}
\end{equation}
where $\text{div}(v) = \nabla_\mu v^\mu$ is the divergence of $v$.  (We show that $\mathcal L_v(i_v \epsilon) = \text{div}(v) \. i_v\epsilon$ in Appendix \ref{lieproof}.)  This demonstrates that the vector volume element contribution from each integral curve is independent of position along the curve if and only if vector field be divergence-free. This definition emphasizes that the total vector field volume can be interpreted as the Riemann sum of contributions $\delta \mathcal V_v$ from each individual integral curve of $v$, and the result \eqref{liev} shows that $\delta \mathcal V_v$ is constant along each integral curve. This is the reason why $\mathcal V_{v}$ is independent of the choice of hypersurface.

\section{Lie Derivative Proof} \label{lieproof}
We wish to prove that
\begin{equation}\label{bb}
\mathcal L_v (i_v \epsilon) = \text{div}(v) i_v \epsilon
\end{equation}
as required by (\ref{liev}). We start with Cartan's identity which states that
\begin{equation}
\mathcal L_X \omega = \mathrm d (i_X \omega) + i_X \mathrm d \omega
\end{equation}
for vector $X$, differential form $\omega$, and exterior derivative $\mathrm d$. Note that $i_X^2 = 0$ and $\mathrm d^2 = 0$. We have
\begin{equation}
\mathcal L_v (i_v \epsilon) = \mathrm d(i_v (i_v \epsilon) )+ i_v \mathrm d (i_v \epsilon)
\end{equation}
with $\mathrm d (i_v (i_v \epsilon)) = \mathrm d (i_v^2 \epsilon) = 0$ since $i_v^2 \omega = 0$ for any differential form $\omega$.  Further, we can write $\mathrm d(i_v \epsilon) = \mathcal
L_v \epsilon - i_v \mathrm d \epsilon$ by applying Cartan's identity again.  The $i_v \mathrm d \epsilon$ term becomes zero when the interior product $i_v$ is taken with it.  As a result, we can write
\begin{equation}
\mathcal L_v (i_v \epsilon) = i_v \mathcal L_v \epsilon.
\end{equation}
Now from the definition of divergence we have $\mathcal L_v \epsilon = \text{div}(v) \epsilon$ and so
\begin{equation}
\mathcal L_v (i_v \epsilon) = i_v (\text{div}(v) \epsilon).
\end{equation}
Since the interior product is the contraction of a form with the vector field, the scalar $\text{div}(v)$ can be brought outside the interior product and we arrive at (\ref{bb}) as required.


\begin{thebibliography}{}
\bibitem{notation} Throughout this paper we use Greek indices to go from 0 to $D-1$, where $D$ is the dimensionality of the spacetime; Roman indices usually go from $1$ to $D-1$ but should be clear in context.
\bibitem{cveticfootnote} Cveti\v{c} et al.~define their integral differently, by letting $r^2$ take on negative values and taking the integral from $r^2 =-a^2$ up to $r^2 = r_+^2$, but the result of the volume integral is the same, since the integral has a term $r(r^2+a^2)$ which is zero for both $r = 0$ and $r^2 = -a^2$.
\end{thebibliography}
\end{document}